\documentclass{aastex631}

\shorttitle{GRB 211024B}
\shortauthors{Fu et al.}
\graphicspath{{./}{figures/}}
\usepackage{amsmath}

\begin{document}

\title{GRB 211024B: an ultra-long GRB powered by magnetar}

\correspondingauthor{Dong Xu, Wei-Hua Lei}
\email{dxu@nao.cas.cn, leiwh@hust.edu.cn}

\author[0009-0002-7730-3985]{Shao-Yu Fu}
\affiliation{Key Laboratory of Space Astronomy and Technology, National Astronomical Observatories, Chinese Academy of Sciences, Beijing, 100101, China}
\affiliation{School of Astronomy and Space Science, University of Chinese Academy of Sciences, Chinese Academy of Sciences, Beijing 100049, China}

\author[0000-0003-3257-9435]{Dong Xu}
\affiliation{Key Laboratory of Space Astronomy and Technology, National Astronomical Observatories, Chinese Academy of Sciences, Beijing, 100101, China}

\author[0000-0003-3440-1526]{Wei-Hua Lei}
\affiliation{Department of Astronomy, School of Physics, Huazhong University of Science and Technology, Wuhan, 430074, China}

\author[0000-0001-7717-5085]{Antonio~de~Ugarte~Postigo}
\affiliation{Universit\'{e} de la C\^ote d'Azur, Observatoire de la C\^ote d'Azur, CNRS, Artemis, 06304 Nice, France}
\affiliation{Aix Marseille Univ, CNRS, LAM Marseille, France}

\author[0000-0002-7517-326X]{Daniele B. Malesani}
\affiliation{Niels Bohr Institute, University of Copenhagen, Jagtvej 155, 2200, Copenhagen N, Denmark}
\affiliation{Department of Astrophysics/IMAPP, Radboud University, PO Box 9010, 6500 GL, The Netherlands}

\author[0000-0003-2902-3583]{David Alexander Kann}
\affiliation{Instituto de Astrof\'isica de Andaluc\'ia, Glorieta de la Astronom\'ia s/n, 18008 Granada, Spain.}
\affiliation{Hessian Research Cluster ELEMENTS, Giersch Science Center, Max-von-Laue-Strasse 12, Goethe University Frankfurt, Campus Riedberg, 60438 Frankfurt am Main, Germany}

\author{Páll Jakobsson}
\affiliation{Centre for Astrophysics and Cosmology, Science Institute, University of Iceland, Dunhagi 5, 107 Reykjav\' ik, Iceland}

\author[0000-0002-8149-8298]{Johan P. U. Fynbo}
\affiliation{The Cosmic Dawn Centre (DAWN)}
\affiliation{Niels Bohr Institute, University of Copenhagen, Jagtvej 155, 2200, Copenhagen N, Denmark}

\author{Elisabetta Maiorano}
\author{Andrea Rossi}
\affiliation{INAF–Osservatorio di Astroﬁsica e Scienza dello Spazio, via Piero Gobetti 93/3, I-40129 Bologna, Italy}

\author{Diego Paris}
\affiliation{INAF–Osservatorio Astronomico di Roma, via Frascati 33, I-00040 Monte Porzio Catone, Italy}

\author{Xing Liu}
\author{Shuai-Qing Jiang}
\author{Tian-Hua Lu}
\author{Jie An}
\affiliation{Key Laboratory of Space Astronomy and Technology, National Astronomical Observatories, Chinese Academy of Sciences, Beijing, 100101, China}
\affiliation{School of Astronomy and Space Science, University of Chinese Academy of Sciences, Chinese Academy of Sciences, Beijing 100049, China}

\author[0000-0002-9022-1928]{Zi-Pei Zhu}
\affiliation{Key Laboratory of Space Astronomy and Technology, National Astronomical Observatories, Chinese Academy of Sciences, Beijing, 100101, China}

\author{Xing Gao}
\affiliation{Xinjiang Astronomical Observatory, Chinese Academy of Sciences, Urumqi,  Xinjiang 830011, China}

\author{Jian-Yan Wei}
\affiliation{Key Laboratory of Space Astronomy and Technology, National Astronomical Observatories, Chinese Academy of Sciences, Beijing, 100101, China}
\affiliation{School of Astronomy and Space Science, University of Chinese Academy of Sciences, Chinese Academy of Sciences, Beijing 100049, China}

\begin{abstract}

Ultra-long gamma-ray bursts (ULGRBs) are characterized by exceptionally long-duration central engine activities, with characteristic timescales exceeding 1000 seconds. We present ground-based optical afterglow observations of the ultra-long gamma-ray burst GRB 211024B, detected by \textit{Swift}. Its X-ray light curve exhibits a characteristic ``internal plateau" with a shallow decay phase lasting approximately $\sim 15$\,ks, followed by a steep decline ($\alpha_{\rm drop}\sim-7.5$). Moreover, the early optical emission predicted by the late r-band optical afterglow is significantly higher than the observed value, indicating an external shock with energy injection. To explain these observations, we propose a magnetar central engine model. The magnetar collapse into a black hole due to spin-down or hyperaccretion, leading to the observed steep break in the X-ray light curve. The afterglow model fitting reveals that the afterglow injection luminosity varies with different assumptions of the circumburst medium density, implying different potential energy sources. For the interstellar medium (ISM) case with a fixed injection end time, the energy may originate from the magnetar's dipole radiation. However, in other scenarios, relativistic jets produced by the magnetar/black hole system could be the primary energy source. 

\end{abstract}

\keywords{\href{http://astrothesaurus.org/uat/629}{Gamma-ray bursts (629)}}

\section{Introduction} \label{sec:intro}
Gamma-ray bursts (GRBs) are the most energetic explosions in the universe, releasing immense amounts of energy peak in the high-energy gamma-ray band. Traditionally, GRBs have been classified based on their gamma-ray emission duration, with long-GRBs (LGRBs) lasting longer than 2 seconds ($T_{90} > 2$\,s) and short-GRBs (SGRBs) lasting less than 2 seconds ($T_{90} < 2$\,s) \citep{1993ApJ...413L.101K}. The observed association between LGRBs and Type Ic supernovae strongly suggests that at least some LGRBs arise from the collapse of massive stars \citep{2006ARA&A..44..507W}. Conversely, the detection of a gravitational wave envet GW 170817 coincident with GRB 170817A confirmed the theoretical prediction that at least some SGRBs originate from the merger of neutron stars \citep{2017PhRvL.119p1101A, 2017Natur.551...64A, 2017Sci...358.1556C, 2017ApJ...848L..17C, 2017Sci...358.1559K, 2017ApJ...848L..16S}.

In the \textit{Swift} era, the detection efficiency of GRBs has been greatly improved, and we have a clear understanding of the morphological features of X-ray light curves \citep{2006ApJ...642..354Z}. In addition to the power-law decay predicted by the external shock theory, there are also features such as flares and plateaus, which suggest the presence of an active central engine at the burst center. In recent years, a class of GRBs with extremely long durations has been discovered, known as ultra-long GRBs (ULGRBs), such as GRB 060607A \citep{2008MNRAS.385..453Z}, GRB 091024A \citep{2013ApJ...778...54V}, GRB 101225A \citep{2014ApJ...781...13L, 2012Natur.482R.120T}, GRB 111209A \citep{2013ApJ...766...30G, 2013ApJ...779...66S, 2018A&A...617A.122K}, GRB 121027A \citep{2014ApJ...781...13L, 2013arXiv1302.4876P, 2013ApJ...767L..36W}, GRB 130925A \citep{2014MNRAS.444..250E} and GRB 220627A \citep{2023A&A...677A..32D}. Some of these GRBs have gamma-ray emission exceeding 1000\,s, while others exhibit long-term active central engine features in their X-ray light curves. \cite{2014ApJ...781...13L} (L14) studied three ULGRBs and found that their host galaxies are faint, compact, and highly star-forming dwarf galaxies. They also proposed a criterion for identifying ULGRBs: high-energy emission lasting up to $\sim 10^4$\,s; rapid decay after plateau phase, which decay index consistent with high-latitude emission ($\alpha_{\rm drop}\le-3$) and rapid variations (flares or dips) in plateau phase. \cite{2014ApJ...787...66Z} studied the Swift GRBs sample and estimated the burst duration  $t_{\rm burst}$ of the sample. They found that $t_{\rm burst}$ is significantly higher than $T_{90}$, and even 11.5\% of GRBs have $t_{\rm burst} > 10^4$\,s, indicating that the active time of the GRB central engine is longer than it appears.

Due to the longer duration of ULGRBs, blue supergiant stars (BSGs) have been proposed as possible progenitor stars of ULGRBs, because their radius are much larger than those of the progenitor,  Wolf-Rayet (WR) stars, of LGRBs \citep{2001ApJ...556L..37M, 2013ApJ...778...67N}. \cite{2018ApJ...859...48P} performed an end-to-end simulation and found that a BSG explosion can produce an accretion disk with an accretion timescale of $\sim 10^4$\,s. The fragmentation of massive star envelope \citep{2005ApJ...630L.113K} or the accretion disk \citep{2006ApJ...636L..29P} can produce the flares we see in X-ray light curves. Some have proposed that active newborn magnetars can produce such long-term radiation, and will exhibit a plateau seen in the X-ray light curve of some GRBs \citep{2007ApJ...665..599T, 2007ApJ...670..565L, 2014ApJ...785...74L}.

In this paper, we present optical observations of the ultra-long gamma-ray burst GRB 211024B detected by the \textit{Swift}. Its X-ray light curve shows a typical ``internal plateau" feature. Notably, the early-time optical flux is observed to be lower than the predictions based on the extrapolation from late-time afterglow. We propose that these observational features arises from continued activity of the new-born magnetar. In Section 2, we provides a description of the observation and data reduction. In Section 3, we introduce our model and fitting procedure for GRB 211024B. The discussion and and conclusion are presented in Section 4 and Section 5, respectively. Throughout this paper, we adopt a conventional cosmological model with the following parameter values: $H_0 = 69.6$ km s$^{-1}$ Mpc$^{-1}$, $\Omega_M = 0.286$, and $\Omega_{\Lambda} = 0.714$ \citep{2014ApJ...794..135B}.

\section{Observation and Data Reduction} \label{sec:obser}

\subsection{Space Obsevation}
GRB 211024B was discovered by \textit{Swift} Burst Alert Telescope (BAT) on 2021 October 24 at 22:20:36 UT. The prompt emission shows a single-peak structure with a duration of about 70 seconds \citep{2021GCN.30980....1G}. Owing to the conflict of preplanned observation, XRT and UVOT were unable to slew to the target at the first time. Late time light curve of prompt emission observed by BAT shows that the duration of GRB 211024B longer than $\sim 600$s.

XRT begins observation of GRB 211024B at $\sim 5.6$\,ks after the BAT trigger and gives enhanced XRT position RA (J2000) = 10:18:51.04, Dec(J2000) = 24:34:05.7, with an uncertainty of 1.9 arcsec \citep{2021GCN.30994....1E}. The XRT light curve shows a plateau-like structure until 13\,ks followed by a steep decay with index $\sim 5$. The PC mode spectrum can be fitted by absorbed power-law with photon spectral index $\Gamma = 2.08_{-0.10}^{+0.11}$ and absorption column is  $9.7^{+2.7}_{-2.5} \times 10^{20}$ cm$^{-2}$ \citep{2021GCN.30994....1E}.

Optical emission from GRB 211024B was detected by UVOT at $\sim 5.6$\,ks after the BAT trigger with U-band magnitude $18.8 \pm 0.2$. We download \textit{Swift}/UVOT data from the UK \textit{Swift} Science Data Centre\footnote{\url{https://www.swift.ac.uk/swift\_portal/}}and analyse it with the \texttt{uvotproduct} pipeline of the HEAsoft software version 6.31.

\subsection{Ground-base Photometry Observation}
The initial observation automatically started at 22:23:13 UT on 2021-10-24, i.e., 157 s after the BAT trigger, using the NEXT-0.6m optical telescope located at Nanshan, Xinjiang, China. We obtaine a series of frames in the Sloan r-filter and covered prompt emission phase for at least the initial 600s. However, there is no obvious correlation between optical and prompt emission light curve. Subsequent photometry follow-up observation were carried out using the 2.56 m Nordic Optical Telescope (NOT; Roque de los Muchachos observatory, La Palma, Spain) equipped with Alhambra Faint Object Spectrograph and Camera (ALFOSC) from October 26 to December 29, 2023. In addition, we observed the burst in the Sloan \textit{r} and \textit{z} bands with Large Binocular Telescope (LBT) on November 7, 2021 (about 13 days after the burst), and performed \textit{griz} multi-color photometry of its host galaxy on April 25, 2022 (about 182 days after the burst).

All optical data are reduced by standard procedures with the Image Reduction and Analysis Facility (IRAF) v2.16 \citep{1986SPIE..627..733T}. For each frame of NEXT, bias and dark subtraction, flat-fielding, and cosmic ray removal were implemented. At each epoch of NOT observation, bias subtraction, flat-fielding, and image combination were performed. Aperture photometry was conducted on each frames with an aperture radius of 2.5 times full width at half maximum (FWHM) Flux calibration was calibrated by referencing nearby Pan-STARRS1 field stars \citep{2016arXiv161205560C}.

\begin{figure}[ht!]
\plotone{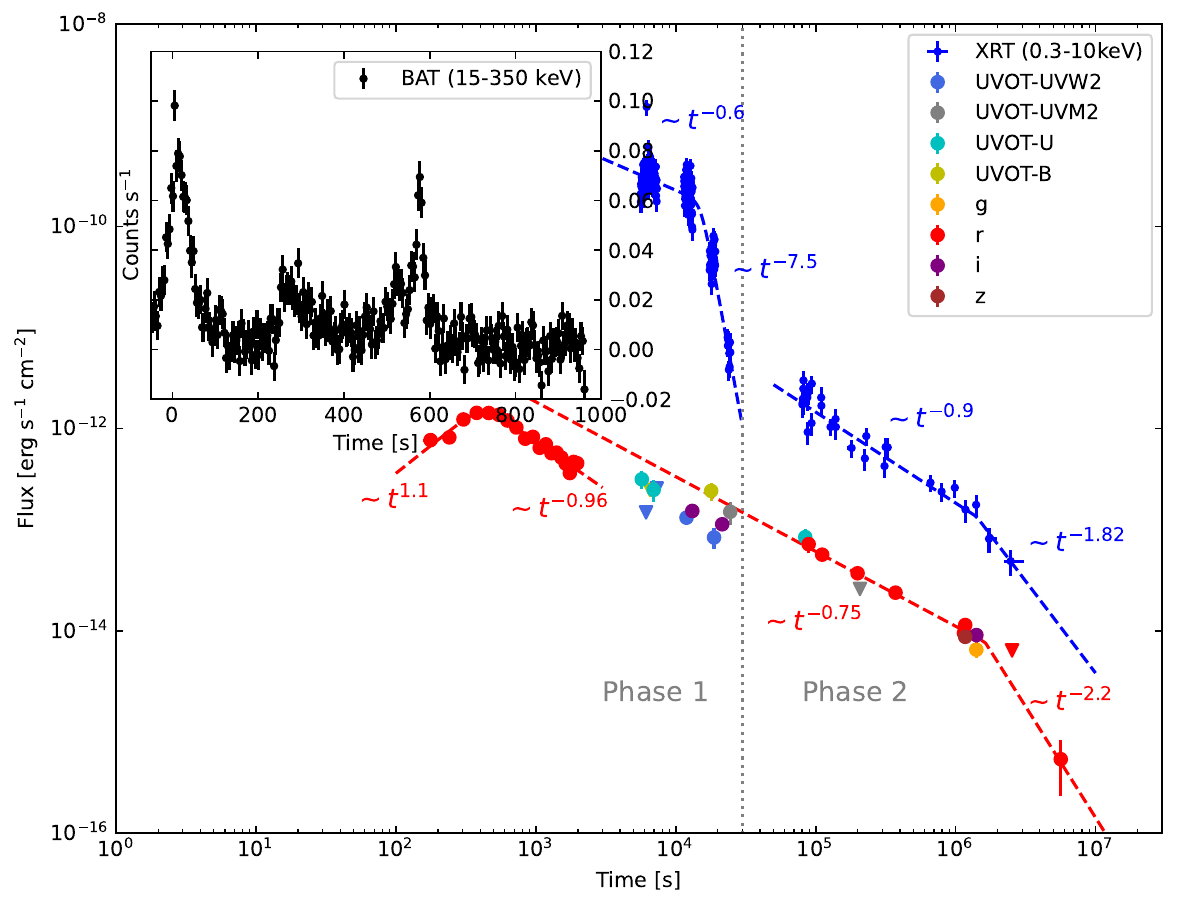}
\caption{Multi-band prompt emission and afterglow light curve of GRB 211024B in logarithmic timescale. The inset shows the \textit{Swift}/BAT light curve in linear timescale. Inverted triangles represent $3-\sigma$ upper limits. Dashed lines depict the best-fit smoothly broken power law (SBPL) models to the light curves. Note that the data have been subtracted for the host galaxy flux contribution and no Galactic extinction correction has been made. \label{fig.lc1}}
\end{figure}

In addition to our own observations, we collated observation results from the Gamma-ray Coordinates Network \citep[GCN,][]{2021GCN.30995....1K, 2021GCN.31028....1J}. All photometry results are reported in Table \ref{tb.phot}, while the corresponding light curves shown in Figure \ref{fig.lc1}.

\begin{deluxetable*}{cccccc}
\tablecaption{Optical photometry observation of GRB 211024B}
\label{tb.phot}
\tablewidth{0pt}
\tablehead{
\colhead{$T-T_0$ (days)} & \colhead{Band} & \colhead{Mag (AB)} & \colhead{Err} & \colhead{Ins.} & \colhead{References}
}
%\decimalcolnumbers
\startdata
0.0020  & r & 18.40  & 0.13  & NEXT & this work \\ 
0.0028  & r & 18.33  & 0.12  & NEXT & this work \\ 
0.0035  & r & 17.89  & 0.09  & NEXT & this work \\ 
0.0044  & r & 17.72  & 0.06  & NEXT & this work \\ 
0.0053  & r & 17.73  & 0.06  & NEXT & this work \\ 
0.0063  & r & 17.77  & 0.06  & NEXT & this work \\ 
0.0073  & r & 17.91  & 0.07  & NEXT & this work \\ 
0.0084  & r & 18.08  & 0.06  & NEXT & this work \\ 
0.0097  & r & 18.36  & 0.08  & NEXT & this work \\ 
0.0110  & r & 18.32  & 0.08  & NEXT & this work \\ 
0.0123  & r & 18.59  & 0.10  & NEXT & this work \\ 
0.0137  & r & 18.50  & 0.09  & NEXT & this work \\ 
0.0150  & r & 18.72  & 0.11  & NEXT & this work \\ 
0.0163  & r & 18.71  & 0.11  & NEXT & this work \\ 
0.0176  & r & 18.83  & 0.12  & NEXT & this work \\ 
0.0189  & r & 18.98  & 0.14  & NEXT & this work \\ 
0.0203  & r & 19.21  & 0.17  & NEXT & this work \\ 
0.0216  & r & 18.94  & 0.13  & NEXT & this work \\ 
0.0229  & r & 18.97  & 0.14  & NEXT & this work \\ 
1.0308  & r & 20.97 & 0.23 & NEXT & this work \\ 
1.2880  & r & 21.23 & 0.05 & NOT/ALFOSC & this work \\ 
2.3060  & r & 21.69 & 0.06 & NOT/ALFOSC & this work \\ 
4.3070  & r & 22.17 & 0.07 & NOT/ALFOSC & this work \\ 
13.279  & r & 23.17 & 0.19 & NOT/ALFOSC & this work \\ 
29.323  & r & \textgreater23.6 &  & NOT/ALFOSC & this work \\ 
65.312  & r & 25.08 & 0.29 & NOT/ALFOSC & this work \\
16.291  & i & 23.01 & 0.21 & NOT/ALFOSC & this work \\
16.272  & g & 23.87 & 0.2 & NOT/ALFOSC & this work \\ 
13.561  & r & 22.97 & 0.03 & LBT & this work \\
13.561  & z & 22.86 & 0.07 & LBT & this work \\
182.39  & g & 25.74 & 0.31 & LBT & this work \\
182.39  & r & 25.51 & 0.39 & LBT & this work \\
182.39  & i & \textgreater25.0 &   & LBT & this work \\
182.39  & z & \textgreater24.5 &   & LBT & this work \\
0.2485  & i & 20.27 & 0.05 & CAHA/CAFOS & \cite{2021GCN.30995....1K} \\ 
0.1520  & i & 19.94 & 0.11 & D50 & \cite{2021GCN.31028....1J} \\ 
0.0757  & B & \textgreater20.01 &  & \textit{Swift}/UVOT & this work \\ 
0.2073  & B & 20.03 & 0.24 & \textit{Swift}/UVOT & this work \\ 
0.0662  & U & 19.99 & 0.24 & \textit{Swift}/UVOT & this work \\ 
0.0805  & U & 20.25 & 0.31 & \textit{Swift}/UVOT & this work \\ 
0.9751  & U & 21.42 & 0.23 & \textit{Swift}/UVOT & this work \\ 
22.568  & U & \textgreater22.78 &  & \textit{Swift}/UVOT & this work \\ 
0.0710  & UVW2 & \textgreater21.34 &  & \textit{Swift}/UVOT & this work \\ 
0.0846  & UVW2 & \textgreater20.75 &  & \textit{Swift}/UVOT & this work \\ 
0.1382  & UVW2 & 21.47 & 0.19 & \textit{Swift}/UVOT & this work \\ 
0.2174  & UVW2 & 21.96 & 0.29 & \textit{Swift}/UVOT & this work \\ 
0.2836  & UVM2 & 21.26 & 0.32 & \textit{Swift}/UVOT & this work \\ 
2.3981  & UVM2 & \textgreater23.17 &  & \textit{Swift}/UVOT & this work \\ 
\enddata
\tablecomments{The photometric data presented in this table have not been corrected for either Galactic or host galaxy extinction.}
\end{deluxetable*}

\subsection{Spectroscopic Observation}
We obtained a spectrum of the GRB 211024B host galaxy on 27 March 2022, 153 days after the GRB, using the X-shooter spectrograph mounted on Unit Telescope 3 of the European Southern Observatory's Very Large Telescope. The observation consisted of 4 exposures of 1200 seconds each in the UVB and VIS arms, and 8 exposures of 600 seconds each in the NIR arm, covering a wavelength range of 3000 to 21000\,\AA. A portion of the K-band was intentionally blocked by the JH-slit to improve the efficiency of the J and H bands.

\begin{figure}[ht!]
\plotone{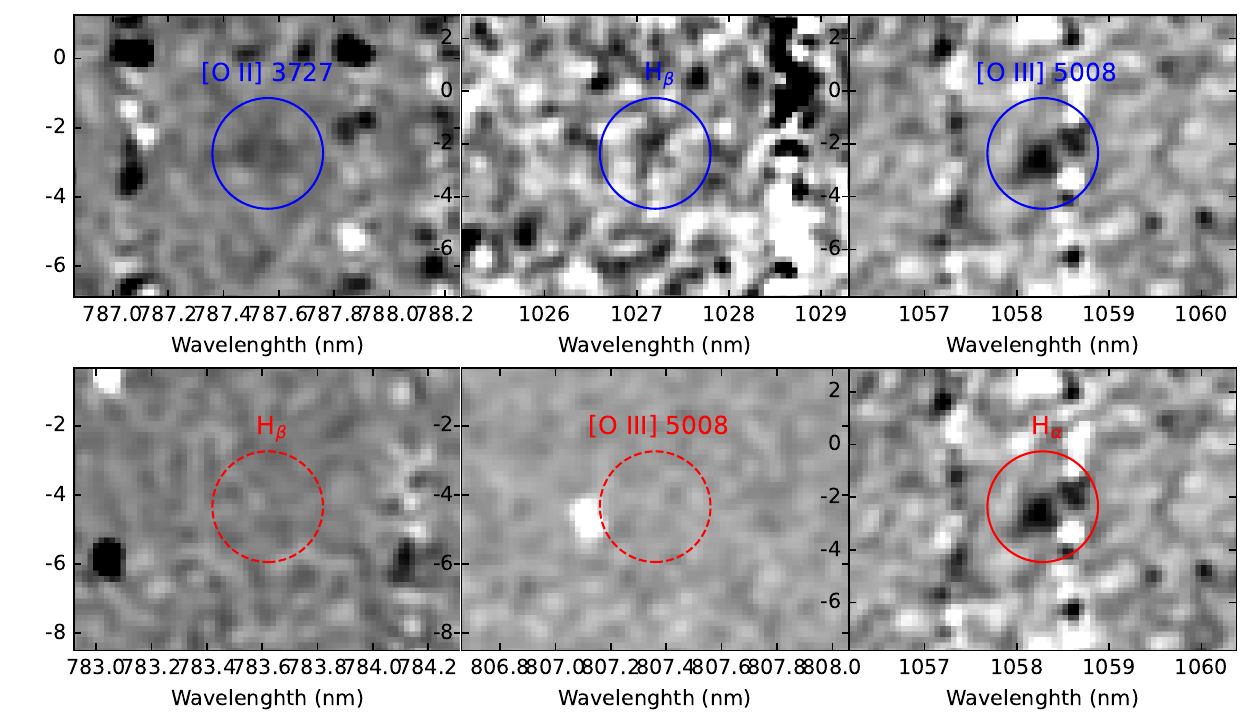}
\caption{Cut-outs of VLT/X-shooter 2D-spectrum of the host galaxy. The emission line at 10582 \AA\ is [OIII] (top right panel), we can derive redshift of $z = 1.113$, which is supported by the weak signals of [OII] seen at 7875 \AA\ (top left panel) and H$_{\beta}$ at 10271 \AA\ (top middle panel). However, if we assume that the emission line at 10582 \AA\ is H$_{\alpha}$ (bottom right panel), implying a corresponding redshift of $z = 0.612$, we should expect to see [OIII] at 8073 \AA\ (bottom middle panel) and possibly H$_{\beta}$ at 7836 \AA\ (bottom left panel), but no evidence of emission lines could be found in the expected region of 2D-spectrum. \label{fig.2dspec}}
\end{figure}

The host galaxy continuum is not visible in the spectrum, and only possible weak emission line features are detected at 7875, 10271, and 10582 \AA. We tentatively identify the emission line feature at 10582 \AA\ as [OIII] 5008 \AA, yielding a redshift of z = 1.113. This is supported by the weak emission lines of [OII] 3737 \AA\ at 7875 \AA\ and H$_{\beta}$ at 10271 \AA\ (top panel of Figure \ref{fig.2dspec}). The expected H$_{\alpha}$ at 13868 \AA, unfortunately, falls within a telluric absorption region, resulting in non detection. If the emission line feature at 10582 \AA\ is H$_{\alpha}$, it corresponds to a redshift of z = 0.612, and we would expect to see [OIII] 5008 \AA\ at 8073 \AA\ and possibly H$_{\beta}$ at 7836 \AA. In this case, no emission line features are found in the corresponding two-dimensional spectrum (bottom panel of Figure \ref{fig.2dspec}). Moreover, a redshift of $z\sim0.6$ would be a bit low for such a faint galaxy. Finally, if 10582 \AA\ is [O II], we would clearly resolve the [O II] doublet in the NIR arm of the spectrum. Therefore, we conclude that the redshift of the host galaxy is $z = 1.113 \pm 0.001$.

\section{Modeling the afterglow of GRB 211024B}\label{sec.afterglow}

\subsection{Temporal Analysis}\label{subsec.temporal}
The X-ray light curve can be divided into two phases: a plateau-like phase before 15 ks, followed by a steep decay; and a regular decay phase with a jet-break at 10 days. Among the ground-based optical observations, the r-band data is the most plentiful. It can also be divided into two phases: a bump-like phase before 3\,ks and a regular afterglow decay phase after one day. Therefore, we use the smoothly broken power law (SBPL) to fit the light curves of each phase in these two bands. The definition of the SBPL is as follows\footnote{\url{https://docs.astropy.org/en/stable/api/astropy.modeling.powerlaws.SmoothlyBrokenPowerLaw1D.html\#astropy.modeling.powerlaws.SmoothlyBrokenPowerLaw1D}}:
\begin{equation}
f(t)=A\left(\frac{t}{t_b}\right)^{-\alpha_1}\left\{\frac{1}{2}\left[1+\left(\frac{t}{t_b}\right)^{1/s}\right]\right\}^{(\alpha_1-\alpha_2)s}
\end{equation}
where $A$ is model amplitude at the break time, $t_b$ is the break time, $s$ is smoothness parameter, $\alpha_1$ and $\alpha_2$ are power law index before and after break time, respectively. It should be noted that we do not include the LBT observation of the host galaxy at 182 days in the r-band fitting. The fitting results are presented in Figure \ref{fig.lc1} and Table \ref{tb.index}.

\begin{deluxetable*}{ccccccc}
\tablecaption{Measurements of temporal index}
\label{tb.index}
\tablewidth{0pt}
\label{tb.temp_fit}
\tablehead{
\colhead{Band} & \colhead{Phase} & \colhead{$\alpha_1$} & \colhead{$\alpha_2$} & \colhead{$\log t_b$} & \colhead{$s$} & \colhead{$\chi_r^2 (d.o.f)$}
}
\startdata
X-ray & 1 & $0.60_{-0.02}^{+0.02}$ & $7.50_{-0.35}^{+0.46}$ & $4.18_{-0.01}^{+0.01}$ & $0.06_{-0.03}^{+0.02}$ & 2.15\\
X-ray & 2 & $0.90_{-0.04}^{+0.04}$ & $1.82_{-0.60}^{+0.69}$ & $6.15_{-0.15}^{+0.13}$ & $0.01^*$ & 1.48\\
r & 1 & $-1.10_{-0.37}^{+0.25}$ & $0.96_{-0.07}^{+0.08}$ & $2.62_{-0.04}^{+0.03}$ & $0.12_{-0.06}^{+0.08}$ & 0.79\\
r & 2 & $0.74_{-0.06}^{+0.06}$ & $2.20_{-0.60}^{+0.55}$ & $6.21_{-0.19}^{+0.22}$ & $0.01^*$ & 0.25\\
\enddata
\tablecomments{Fitting results of X-ray and r-band light curve with SBPL model. Parameters marked with asterisks indicate that they are fixed during the fitting. }
\end{deluxetable*}

As shown in Figure \ref{fig.lc1}, the r-band light curve in Phase 1 shows an initial bump with rising power-law index $\alpha_{r,1}\sim 1.1$, peaking time $417$\,s, and decaying power-law index $\alpha_{r,2}\sim -0.96$. We also find a shallow decay following this bump. In Phase 2, the r-band light curve exhibits a typical power-law decay with an index of $\alpha_{r,3}\sim -0.75$, followed by a break at $\sim 1.6\times 10^6$\,s and a subsequent decay with a power-law index of $\alpha_{r,4}\sim -2.2$. On the other hand, the X-ray light curve in Phase 1 shows a shallow decay with a power-law index of $\alpha_{X,1}\sim -0.6$ initially, followed by a break at $\sim 15$\,ks and a steep decay with a slope of $\alpha_{X,2}\sim -7.5$. The X-ray light curve in Phase 2 is generally consistent with the optical r-band, starting with a typical power-law decay with an index of $\alpha_{X,3}\sim -0.9$ and then decaying with an index of $\alpha_{X,4}\sim -1.82$ after a break at $1.4\times 10^6$\,s.
%However, the observed r-band flux is significantly shallower than the extrapolation at the early time.

It is worth noting that the NOT observation at 65 days after the burst may be contaminated by the host galaxy flux. We estimate the host galaxy flux using the LBT observation at 182 days. After subtracting the host galaxy flux contribution, although this leads to a large uncertainty in the afterglow flux, we can still see that the break times as well as the decay index of the r-band and X-ray light curves are consistent. This is likely due to a jet-break effect, which can be used to constrain the half-opening angle of the jet.

\subsection{Theoretical Interpretation}
As shown above, the X-ray light curve of GRB 211024B exhibits a ``plateau'' followed by a sharp decay, then a normal decay and finally a break. The optical light curve shows an initial bump followed by a shallow decay, and then a normal decay followed by a break. In this section, we will show that the initial optical bump and the X-ray/optical normal decay phase are consistent with the standard afterglow model predictions. However, the interpretation for the X-ray plateau and optical shallow decay demand late time central engine activity.

\subsubsection{``Internal Plateau'' in X-ray and Central Engine Activity}
The near-flat plateau ($\alpha_1 \sim 0.6$) followed by a sharp decay ($\alpha_2 \sim 7.5$) in early X-ray afterglow ($t>2\times 10^4$ s) can not be accommodated in any external shock model. Such ``internal plateau'' has to be attributed to the internal dissipation of a central engine \citep{Lv2015,Chen2017}, most likely a supera-massive millisecond magnetar. The abrupt drop can be interpreted as the collapse of the supera-massive magnetar into a black hole \citep{2007ApJ...665..599T, Lv2015, Chen2017}. In this phase, the jet internal dissipation power $L_{\rm dis}$ and the X-ray isotropic luminosity $L_{X,iso}$ is related as 
\begin{equation}
    L_{\rm X,iso} =  L_{\rm dis}/f_{\rm b},
\end{equation}
where $f_{\rm b}=1-\cos \theta_j$ is the beaming factor of the jet.

\subsubsection{Optical Shallow Decay and Energy Injection to Forward Shock}

After the sliding phase, the relativistic blast wave collides with the circumburst medium (CBM), giving rise to long-lived multi-band afterglow emission. 
%A pair shock were generate, namely revers shock (RS) and forward shock (FS). The RS propagate into the shell and an FS propagate into the ISM \citep{Rees1992,Meszaros1997,Sari1998, Sari1999,Zou2005}. 
We consider forward shock (FS) of a relativistic thin shell with energy $E_{\rm K,iso}$, initial Lorentz factor $\Gamma_0$, and opening angle $\theta_{\rm j}$ expanding into the CBM with density $n_0$  \citep{Rees1992,Meszaros1997,Sari1998, Sari1999,Zou2005}. The dynamics of the shell is described by following equation \citep{Huang2000}

\begin{equation}
\frac{dR}{dt} =\beta c \Gamma (\Gamma+\sqrt{\Gamma^2 -1}),    
\end{equation}
\begin{equation}
\frac{dm}{dR} =2\pi R^2 (1-\cos\theta_{\rm j})n_0 m_{\rm p},   
\end{equation}
\begin{equation}\label{eq.dgamma}
\frac{d\Gamma}{dm} = -\frac{\Gamma^2 -1}{M_{\rm ej} + 2\Gamma m} , 
\end{equation}
where $R$ and $t$ are the radius and time of the event in the burst frame, $m$ is the swept-up mass, $M_{\rm ej}=E_{\rm K, iso} (1-\cos\theta_{\rm j})/2(\Gamma_0-1)c^2$ is the ejecta mass, $m_{\rm p}$ is the proton mass, and $\beta=\sqrt{\Gamma^2-1}/\Gamma$. 

%As shown in Figure \ref{tb.phot}, the X-ray light curve exhibits a two-part structure characterized by an initial, slowly declining plateau followed by a very steep decay, which is a typical ``internal plateau" feature. This signature suggests the presence of a central engine continuously outputs energy. Part of this energy $L_{\rm inj}$ is expected to be injected into the blast wave, thereby increasing the observed afterglow luminosity. Such observational characteristics are reflected exactly in the r-band light curve: the early flux estimated based on the the late afterglow is significantly higher than the observed value. 
The initial optical bump and X-ray/optical normal decay will be interpreted by the forward shock with the above dynamics \citep{2014ApJ...788...32W}. 

As discussed in Section 3.2.1,  the ``internal plateau'' in X-ray light curve indicates that the central engine restarts during the plateau phase. Besides the energy $L_{\rm dis}$ dissipated into X-ray emission, some part of the jet energy $L_{\rm inj}$ is expected to be injected into the blast wave, thereby increasing the observed afterglow luminosity. Such observational characteristics are reflected exactly in the shallow decay of the r-band light curve: the early flux estimated based on the the late afterglow is significantly higher than the observed value after the initial bump. Therefore, we consider the energy injection from GRB central engine in this phase, the equation (\ref{eq.dgamma}) should be replace by \citep{Geng2013}
\begin{equation}
\frac{d\Gamma}{dm} = -\frac{\Gamma^2 -1- \frac{1-\beta}{\beta c^3}L_{\rm inj} dR/dm}{M_{\rm ej} + 2\Gamma m} ,    
\end{equation}
During the injection time, i.e. $t_{\rm start}<t<t_{\rm end}$, the injected luminosity is $L_{\rm inj}=L_{\rm inj}^0 (t/t_{\rm start})^{-q}$, where $L_{\rm inj}^0$ is the initial injection power, $q$ is the decay power law index, $t_{\rm start}$ and $t_{\rm end}$ are the start and end time for energy injection. By solving these equations with the initial conditions, one can find the evolution of $\Gamma(t)$ and $R(t)$.

Electrons are believed to be accelerated at the shock front to a power-law distribution $N(\gamma_{\rm e}) \propto \gamma_{\rm e}^{-p}$ during the dynamical evolution of FS. Assuming a constant fraction $\epsilon_{\rm e}$ of the shock energy $e_2=4\Gamma^2 n_1 m_{\rm p} c^2$ is distributed into electrons, this defines the minimum injected electron Lorentz factor,
\begin{equation}
\gamma_{\rm m}=\frac{p-2}{p-1} \epsilon_{\rm e} (\Gamma-1)\frac{m_{\rm p}}{m_{\rm e}}    
\end{equation}
where $m_{\rm e}$ is electron mass. We also assume  a fraction $\epsilon_{\rm B}$ of the shock energy resides in the magnetic field generated downstream of the shock. The comoving magnetic field is given by: 
\begin{equation}
B=(32\pi m_{\rm p} \epsilon_{\rm B} n_0)^{1/2} c.   
\end{equation}
The synchrotron power and characteristic frequency from electron with Lorentz factor $\gamma_{\rm e}$ are
\begin{equation}
P(\gamma_{\rm e})\simeq \frac{4}{3} \sigma_{\rm T} c \Gamma^2 \gamma_{\rm e}^2 \frac{B^2}{8\pi}, \ \ \nu(\gamma_{\rm e}) \simeq \Gamma \gamma_{\rm e}^2 \frac{q_{\rm e} B}{2\pi m_{\rm e}c},  
\end{equation}
%\begin{equation}
%\nu(\gamma_{\rm e}) \simeq \Gamma \gamma_{\rm e}^2 \frac{q_{\rm e} B}{2\pi m_{\rm e}c},
%\end{equation}
where $\sigma_{\rm T}$ is the Thomson cross-section, $q_{\rm e}$ is electron charge. The peak spectra power occurs at $\nu(\gamma_{\rm e})$
\begin{equation}
P_{\nu,{\rm max}} \simeq \frac{P(\gamma_{\rm e})}{\nu(\gamma_{\rm e})}=\frac{m_{\rm e}c^2 \sigma_{\rm T}}{3q_{\rm e}} \Gamma B.   
\end{equation}

One can define the critical electron Lorentz factor $\gamma_{\rm c}$ by setting the electron's lifetime equal to the time $t$,
\begin{equation}
\gamma_{\rm c} = \frac{6\pi m_{\rm e}c}{\Gamma \sigma_{\rm T}B^2 t}.
\end{equation}
When cooling due to synchrotron radiation becomes significant, the shape of the electron distribution should be adjusted for $\gamma_{\rm e} >\gamma_{\rm c}$. Given the radiative cooling and the continuous injection of new accelerated electrons, coupled with the synchronous self-absorption effect, a broken power law spectrum can be obtained. This spectrum is segmented into several sections based on three characteristic frequencies: $\nu_{\rm m}$,  $\nu_{\rm c}$ and $\nu_{\rm a}$, respectively \citep{2013NewAR..57..141G}. The $\nu_{\rm m}$ and $\nu_{\rm c}$ are defined by $\gamma_{\rm m}$ and $\gamma_{\rm c}$ respectively, while $\nu_{\rm a}$ is characterized by synchronous self-absorption. The peak flux density is $F_{\nu,{\rm max}}=N_{\rm e,2} P_{\nu,{\rm max}}/4\pi D^2$, where $N_{\rm e,2}=4\pi R^3 n_0/3$ is the total number of electrons in shocked region of CBM and $D$ is the distance of the source.

% The standard afterglow model can be well described by following parameters: the kinetic energy $E_0$, the initial Lorentz factor $\Gamma_0$, the jet half-opening angle $\theta_{\rm j}$, the viewing angle of observer $\theta_{\rm obs}$, the circumburst medium density $n_{18}$ (the density at $R=10^{18}$\,cm), the electron spectral index $p$, the fractions of shock energy that go to electrons $\epsilon_e$ and magnetic ﬁeld $\epsilon_B$.

We perform multi-wavelength data fitting for GRB\,211024B utilizing the established afterglow model code \texttt{PyFRS}\footnote{\url{https://github.com/leiwh/PyFRS}}  \citep[see also][]{2016ApJ...816...20L, 2023ApJ...948...30Z, Zhou2024}, considering two scenarios for the circumburst medium ($n=n_{18}R^{-k}$): uniform interstellar medium (ISM, $k=0$) and wind-like medium ($k=2$). All data used for fitting have been corrected for Galactic extinction. To explore the parameter space, we utilize the Python-based Markov chain Monte Carlo (MCMC) sampler \texttt{emcee} \citep{emcee}. For the X-ray band, only data acquired after $\sim50 $\,ks (Phase 2) were used in the afterglow fitting. If the energy injection of the afterglow originates from the dipole radiation of a magnetar, the expected end time ($t_{\rm end}$) in the source frame is $t_{end}=t_b/(1+z)\sim 7.1$ ks, where $t_b$ is obtained from fitting of the X-ray light curve in Phase 1. However, if the energy comes from other forms of radiation, the end of injection time becomes uncertain. Therefore, we considered both scenarios in our fitting process, setting the energy injection end time either as a fixed value or as a free parameter. 

\begin{figure*}[ht!]
\gridline{\fig{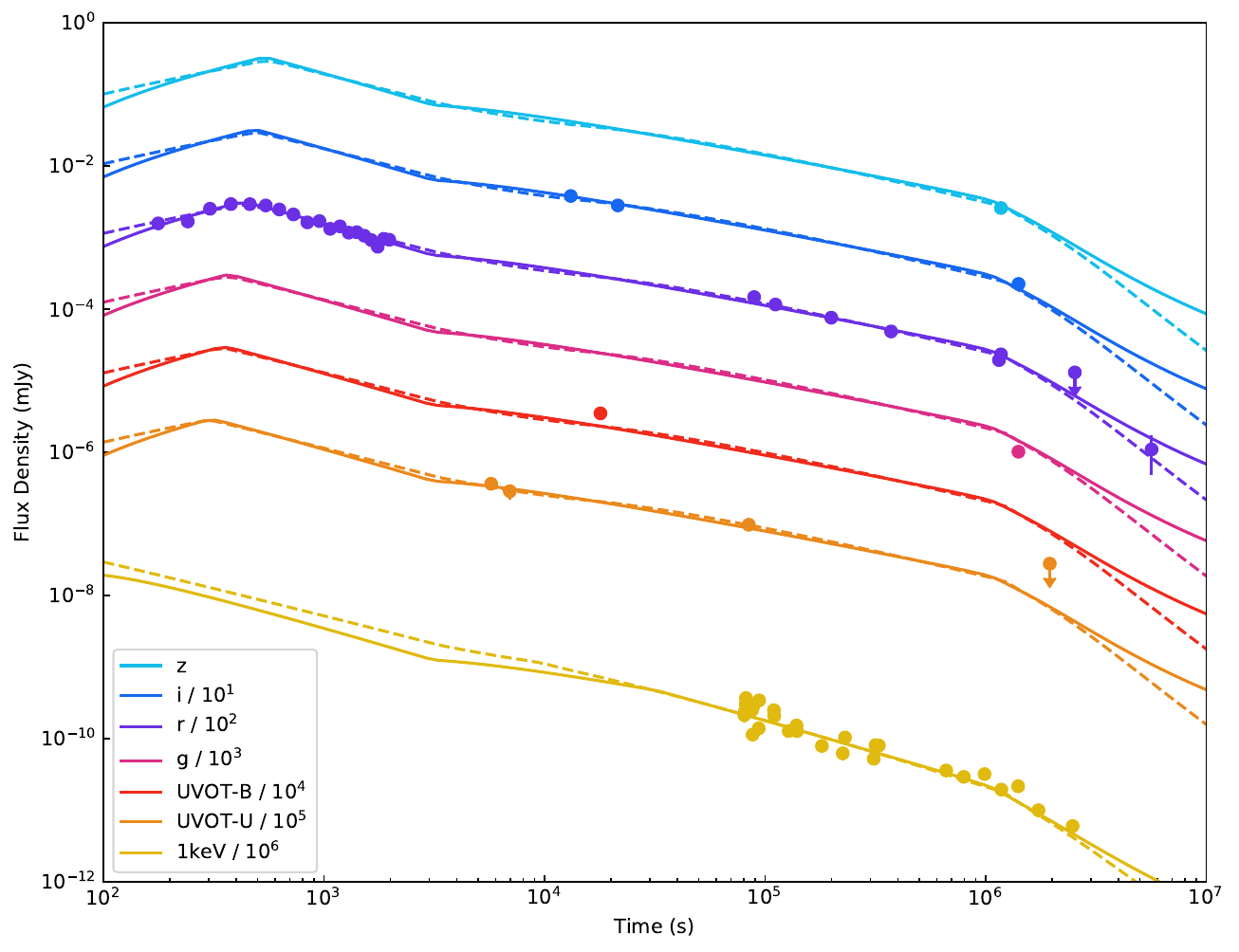}{0.45\textwidth}{(a)}
          \fig{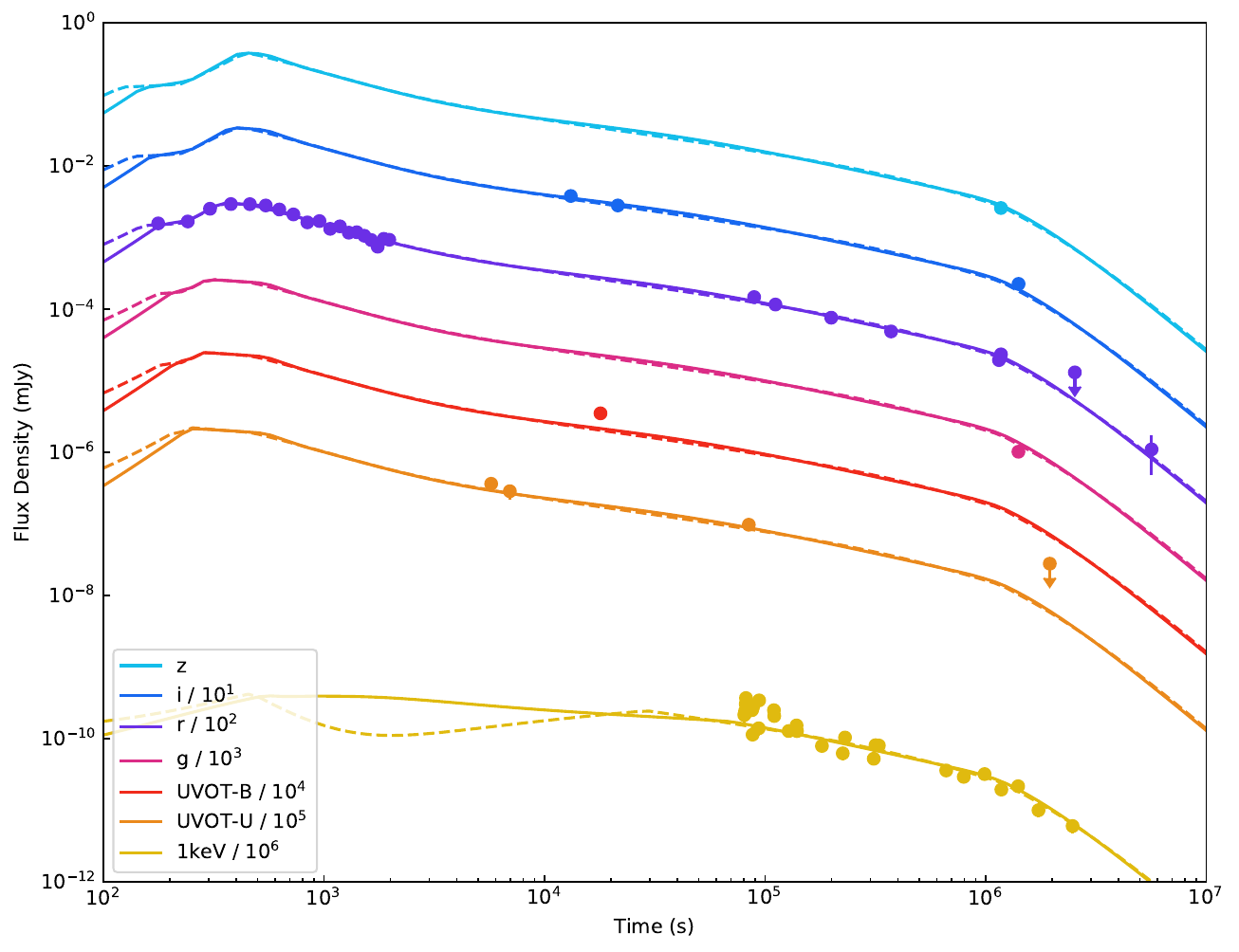}{0.45\textwidth}{(b)}
          }
\caption{The best-fit afterglow light curve of GRB 211024B with energy injection is shown for two scenarios: (a) a uniform interstellar medium (ISM), and (b) a wind-like medium. The solid line represents the case where $t_{\rm end}$ is treated as a free parameter, while the dashed line corresponds to a fixed $t_{\rm end}$. \label{fig.bestfit_lc}}
\end{figure*}

\begin{figure*}[ht!]
% \plotone{AG_corner.pdf}
\gridline{\fig{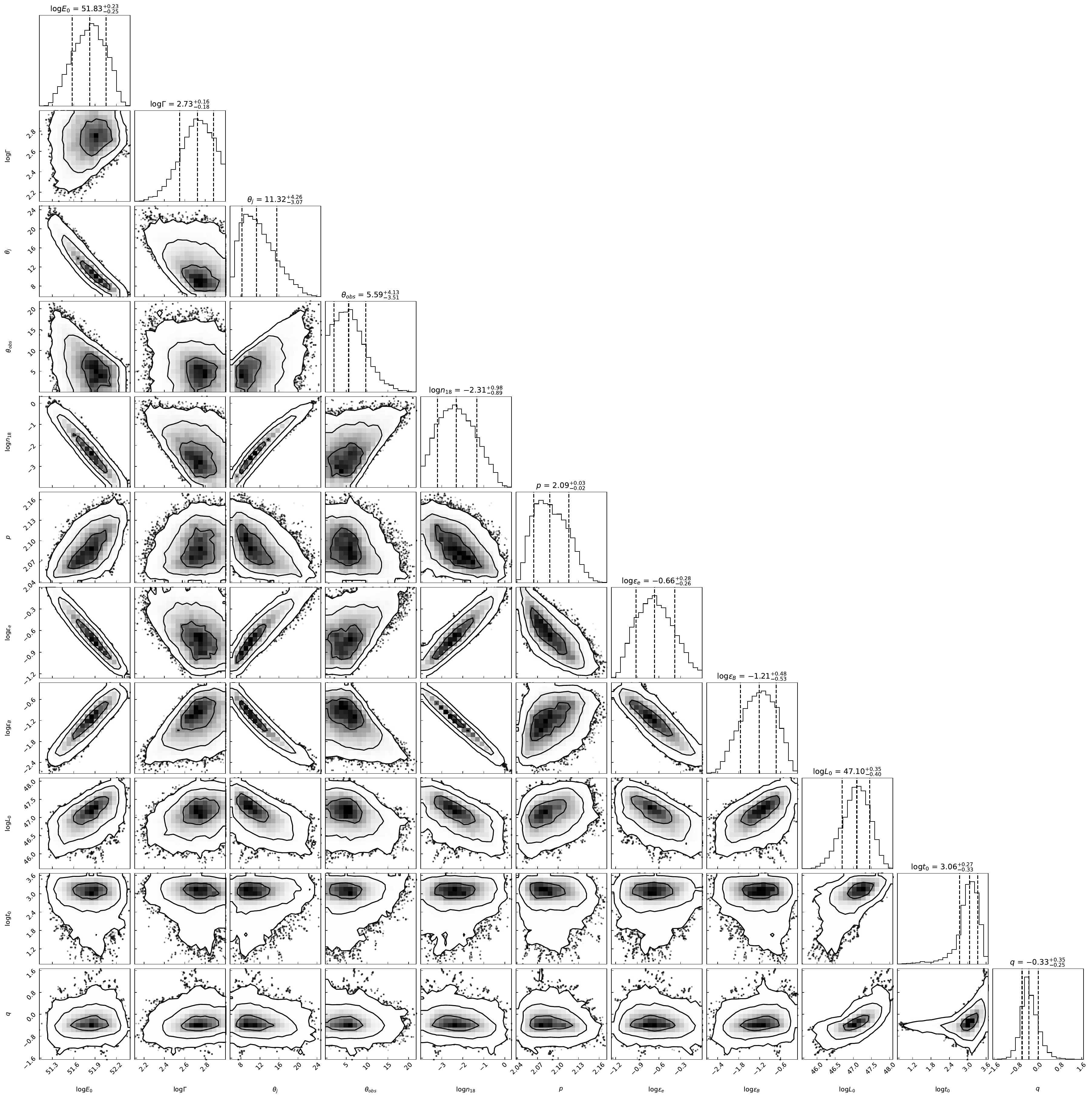}{0.45\textwidth}{(a)}
          \fig{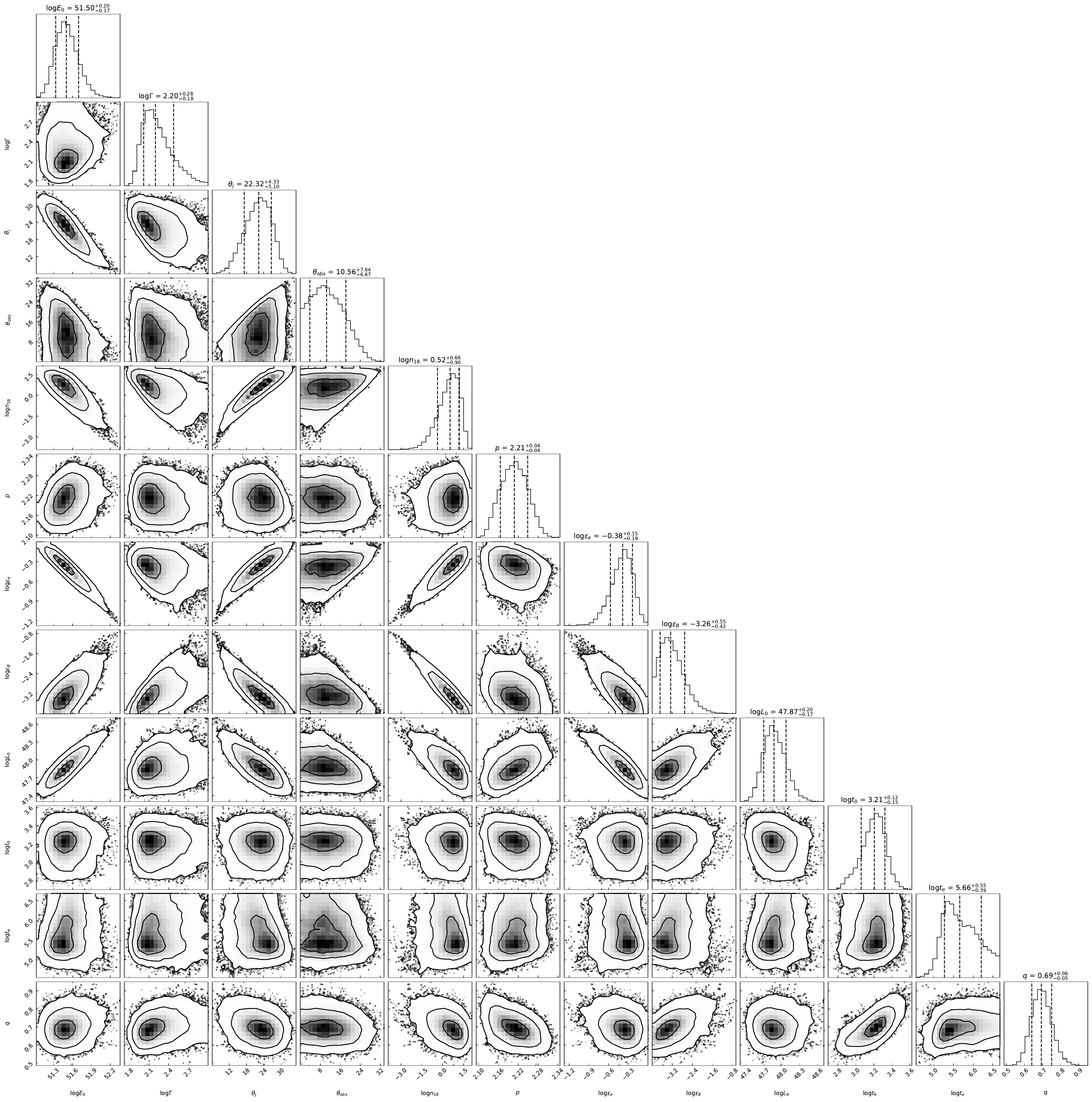}{0.45\textwidth}{(b)}
          }
\gridline{\fig{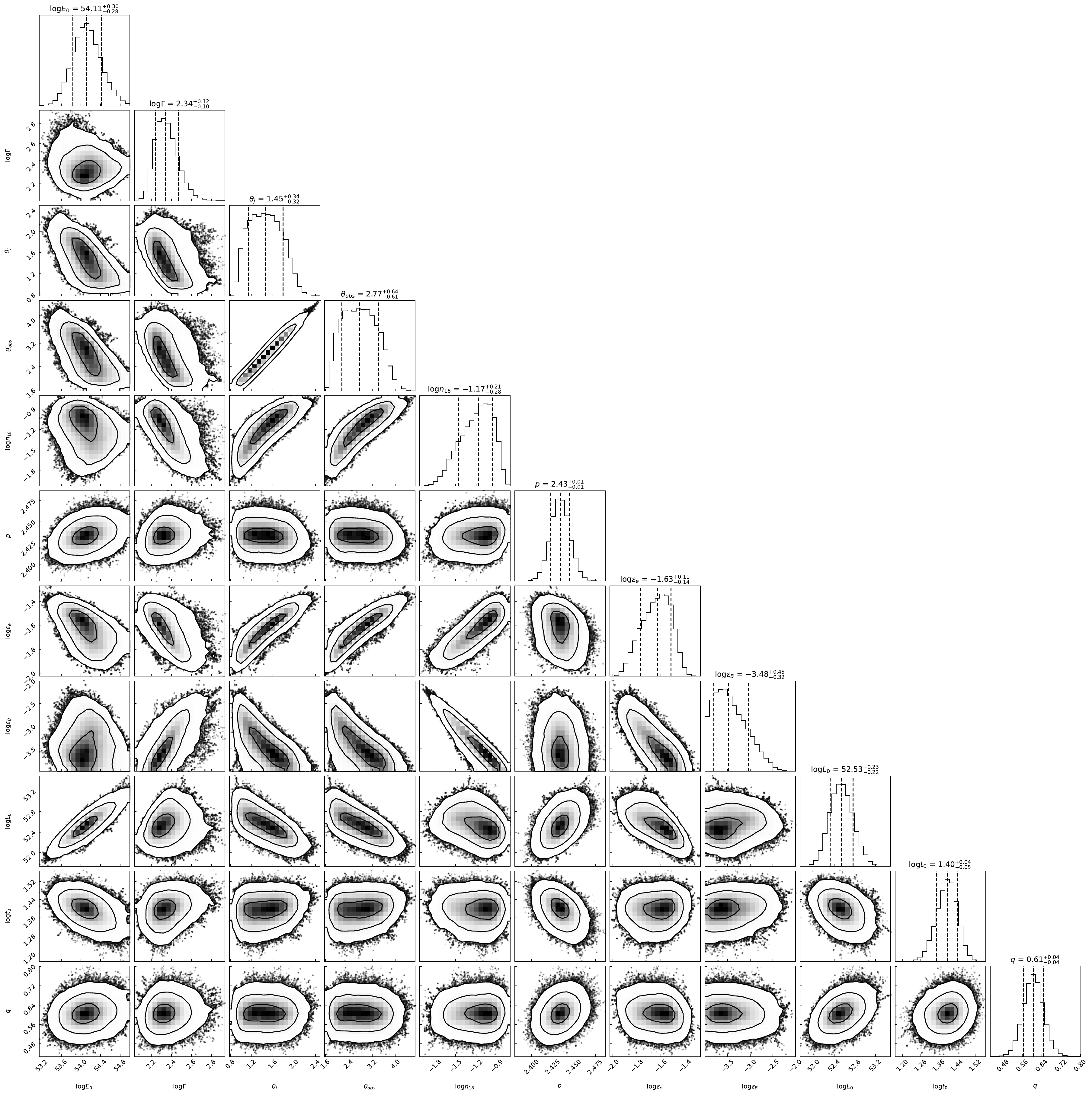}{0.45\textwidth}{(c)}
          \fig{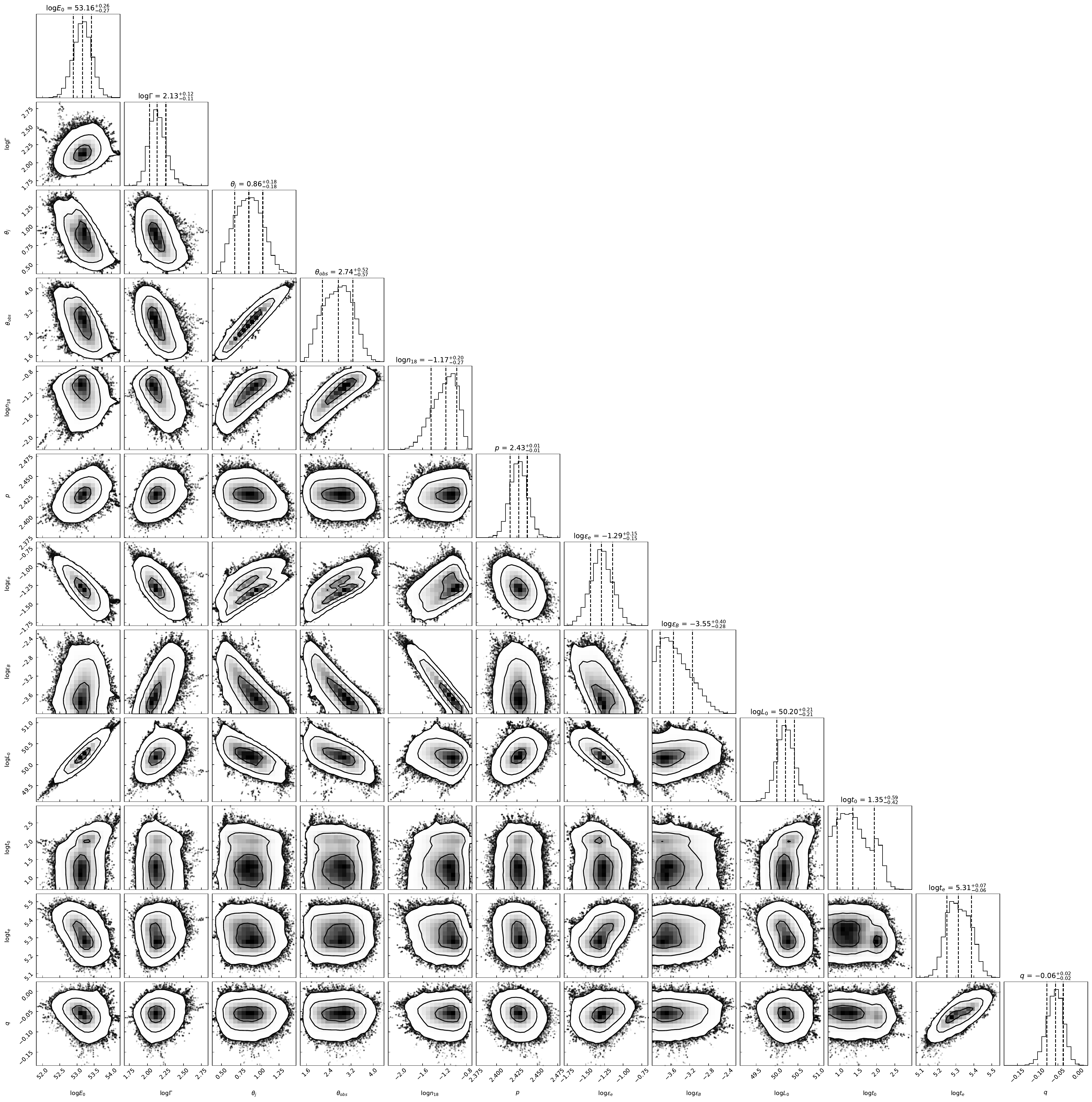}{0.45\textwidth}{(d)}
          }
\caption{Posterior probability distributions of the afterglow parameters of GRB 211024B for different cases: (a) ISM with fix $t_{\rm end}$; (b) ISM with free $t_{\rm end}$; (c) wind-like medium with fix $t_{\rm end}$; (d) wind-like medium with free $t_{\rm end}$. \label{fig.corner}}
\end{figure*}

The MCMC analysis using 120 walkers and 40,000 steps is employed to explore the parameter space, discarding the first 20,000 steps as burn-in. Under this setting, all fits satisfy the condition that the chain length exceeds 80 times the estimated autocorrelation time, and the estimate of the autocorrelation time varies by less than 1\%, indicating that the fits to have converged sufficiently. We employed the Bayesian Information Criterion (BIC) to assess the goodness of fit. The resulting best-fit parameters are presented in Table \ref{tb.AGmodel} and visualized in Figure \ref{fig.bestfit_lc}. The posterior probability distribution of the parameters is shown in Figure \ref{fig.corner}.

Regardless of the scenario, the model fits the observational data well. The Bayesian Information Criterion (BIC) values for different scenarios are not significantly different. For the ISM scenario, the best-fit results generally favor wide jet opening angles ($\theta_j > 10^{\circ}$) and low isotropic kinetic energy ($E_k \sim 10^{51}$ erg). The injection luminosity is around $10^{47}$ erg s$^{-1}$ for both the fixed $t_e$ and free $t_e$ cases, but the former gives a nearly constant luminosity ($q \sim 0$), while the latter shows a gradual decay ($q \sim 0.7$).

The wind-like scenario yields relatively high isotropic kinetic energies ($E_k \sim 10^{53}-10^{54}$ erg) and narrow jet opening angles ($\theta_j \sim 1 $), with injection luminosities reaching $10^{50}-10^{52}$ erg s$^{-1}$. For a magnetar with a typical rotational energy of $E \sim 10^{52} $ erg, such high luminosities are difficult to sustain via magnetic dipole radiation, suggesting that the injection energy in this scenario may originate from other sources such as fall-back accretion-powered jets.

\begin{deluxetable}{cccccccc}
\tabletypesize{\scriptsize}
\tablecaption{Parameters of afterglow modeling}
\label{tb.AGmodel}
\tablewidth{0pt}
\tablehead{
\colhead{Parameters} & \colhead{Unit} & \colhead{Prior Type} & \colhead{Prior} & 
\multicolumn{4}{c}{Results}\\
\cline{5-8}
\colhead{} & \colhead{} & \colhead{} & \colhead{} & \colhead{k=0 (fix $t_{\rm end}$)} & \colhead{k=0} & \colhead{k=2 (fix $t_{\rm end}$)} & \colhead{k=2}
}
\startdata
$E_{\rm k,iso}$  & erg & log-uniform & $[10^{51}, 10^{56}]$ &  $6.76^{+5.26}_{-2.78}\times 10^{51}$ & $3.16^{+1.85}_{-1.02}\times 10^{51}$   & $1.29^{+1.28}_{-0.61}\times 10^{54}$  & $1.45^{+1.18}_{-0.67}\times 10^{53}$ \\
$\Gamma_0$  & 1 & log-uniform & $[1, 1000]$ &  $537^{+239}_{-182}$ & $158^{+144}_{-53.8}$  & $219^{+69.6}_{-45.0}$  & $135^{+42.9}_{-30.2}$ \\
$p$ & uniform & 1 &  $[2.01, 4.0]$ &  $2.09^{+0.03}_{-0.02}$ & $2.21^{+0.04}_{-0.04}$  & $2.43^{+0.01}_{-0.01}$  & $2.43^{+0.01}_{-0.01}$ \\
$n_{18}$$^*$ & cm$^{-3}$ & log-uniform& $[10^{-8}, 10^3]$ & $4.90^{+41.9}_{-4.27}\times 10^{-3}$ & $3.31^{+11.8}_{-2.89}$  & $6.76^{+4.20}_{-3.21}\times 10^{-2}$  & $6.76^{+3.95}_{-3.13}\times 10^{-2}$ \\
$\theta_j$ & degrees & uniform & $[0.01, 45]$ & $11.32^{+4.26}_{-3.07}$  & $22.32^{+4.33}_{-5.10}$   & $1.45^{+0.34}_{-0.32}$  & $0.86^{+0.18}_{-0.18}$ \\
$\theta_{\rm obs}$ & degrees & uniform & $[0.01, 45]$ & $5.59^{+4.13}_{-3.51}$  & $10.56^{+7.64}_{-6.67}$   & $2.77^{+0.64}_{-0.61}$  & $2.74^{+0.52}_{-0.57}$ \\
$\epsilon_e$ & 1 & log-uniform & $[10^{-4}, 1]$ & $2.19^{+1.98}_{-0.99}\times 10^{-1}$ & $4.17^{+1.72}_{-1.48}\times 10^{-1}$  & $2.34^{+0.68}_{-0.65}\times 10^{-2}$  & $5.13^{+2.12}_{-1.50}\times 10^{-2}$ \\
$\epsilon_B$ & 1 & log-uniform & $[10^{-4}, 1]$ & $6.17^{+12.5}_{-4.35}\times 10^{-2}$ & $5.50^{+14.0}_{-3.41}\times 10^{-4}$  & $3.31^{+6.02}_{-1.73}\times 10^{-4}$  & $2.82^{+4.26}_{-1.34}\times 10^{-4}$ \\
\hline
$L_{\rm inj}$  & erg s$^{-1}$ & log-uniform & $[10^{45}, 10^{54}]$ &  $1.26^{+1.56}_{-0.76}\times 10^{47}$ & $7.41^{+4.34}_{-2.40}\times 10^{47}$  & $3.39^{+2.37}_{-1.35}\times 10^{52}$  & $1.58^{+0.99}_{-0.61}\times 10^{50}$ \\
$t_{\rm start}$ & s (source frame) & log-uniform &  $[10^0, 10^5]$ &  $1.15^{+0.99}_{-0.61}\times 10^{3}$ & $1.62^{+0.52}_{-0.47}\times 10^{3}$  & $2.51^{+0.24}_{-0.27}\times 10^{1}$  & $2.24^{+6.47}_{-1.39}\times 10^{1}$ \\
$t_{\rm end}$ & s (source frame) & fixed/log-uniform &  $[10^2,10^7]$ & $7.10\times 10^{3}$ & $4.57^{+11.6}_{-2.71}\times 10^{5}$  & $7.10\times 10^{3}$  & $2.04^{+0.36}_{-0.26}\times 10^{5}$ \\
$q$ & 1 & uniform &  $[-2.0, 2.0]$ &  $-0.33^{+0.35}_{-0.25}$ & $0.69^{+0.06}_{-0.05}$  &  $0.61^{+0.04}_{-0.04}$ & $-0.06^{+0.02}_{-0.02}$ \\
\hline
BIC &  &   &    &  -1794.92 & -1810.95  &  -1762.27 &  -1792.66 \\
\enddata
\tablecomments{* $n_{18}$ is the circumburst medium density at $R=10^{18}$ cm.}
\end{deluxetable}

\section{Discussion}

\subsection{Ultra-long GRB}
The classical classification method of GRBs divides them into short GRBs (SGRBs) and long GRBs (LGRBs) based on the $T_{90}$ criterion of 2 seconds, corresponding to the compact star merger and collapsar origins, respectively. However, there are some exceptions, such as GRB\,060614 \citep{2007ApJ...655L..25Z}, GRB\,160410A \citep{2023MNRAS.520..613A}, GRB\,211211A \citep{2022Natur.612..223R, 2022Natur.612..232Y}, and GRB\,200826A \citep{2022ApJ...932....1R}. The different energy bands of different detectors can lead to some discrepancies in the $T_{90}$ values of GRBs. Moreover, the peak energy of gamma-rays may quickly drop below the energy band of the detector, resulting in a smaller $T_{90}$, even though the central engine is still active. \cite{2014ApJ...787...66Z} studied the \textit{Swift} GRBs sample and found that the active time of the central engine ($t_{\rm burst}$) is generally longer than $T_{90}$. Therefore, ULGRBs, as a unique category of LGRBs, may have different central engines from LGRBs. It is not appropriate to simply use $T_{90}$ as the sole criterion for classification. The criterion proposed by L14 for ULGRBs is a better solution. In addition to requiring a radiation time (consider $\gamma$-ray and X-ray emission both) of more than $10^3$ seconds, it also imposes two additional constraints: the presence of flares or dips in the X-ray plateau phase, and a steep decay following the plateau phase that is consistent with the prediction of high-latitude emission. Plateaus can be classified into two types: ``internal plateau" and ``external plateau". ``Internal plateaus" are directly originate from internal dissipation of central engine. If a steep decay follows the end of the plateau, it characterizes the central engine shutdown timescale. In contrast, ``external plateaus" originates from the long-lasting energy injection of central engine into the external shock. Therefore, the decay index following the plateau is predicted by the external shock model. For ``external plateaus", the end of the plateau may be later than the time of the central engine shutdown, as shown by \cite{2016ApJ...817..152X}. Therefore, the additional criterion of "plateau followed by a steep decay" directly characterizes the duration of the central engine activity.

\begin{figure*}[ht!]
\plotone{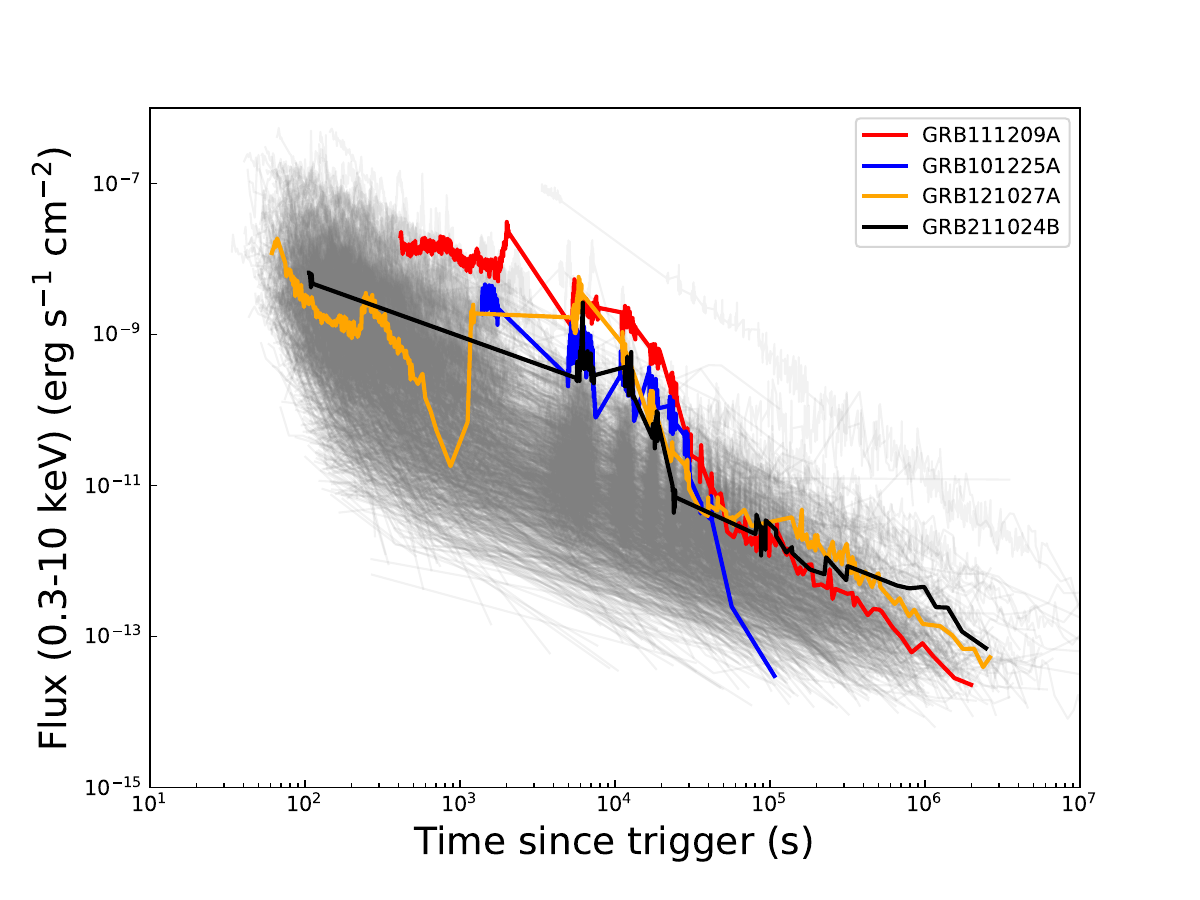}
\caption{X-ray (0.3–10 keV) light curves of \textit{Swift} GRBs. ULGRBs are represented by colored line, ``nomal" GRBs by gray lines, and GRB 211024B by black line. \label{fig.xray_compare}}
\end{figure*}

% \cite{2019ApJS..245....1T} systematically studied 174 GRB X-ray light curves observed by \textit{Swift}, which had measured redshifts and exhibited plateau features. We selected GRBs with a post-plateau decay index greater than -3 (i.e., $\alpha_2>-3$) from their sample, resulting in a total of 162 GRBs. 
Figure \ref{fig.xray_compare} presents the X-ray light curves for 1498 GRBs observed by Swift between December 2004 and May 2023 (gray lines). Three ULGRBs reported by L14 are also plotted as colored lines in Figure \ref{fig.xray_compare}. It is clear that the flux of ULGRBs is significantly higher than that of most "normal" GRBs during the plateau phase ($\sim 10^4$\/s), followed by a steep decay after the plateau. The light curve behavior of GRB 211024B (black line in Figure \ref{fig.xray_compare}) is very similar to that of ULGRBs. Therefore, GRB 211024B is a ULGRB according to the definition of L14.

\subsection{Central Engine Properties}
A supra-massive magnetar may collapse into a black hole after reaching the maximum gravitational mass. The observational correspondence of this prediction is the ``internal X-ray plateau" discovered in some long and short GRBs \citep{2014ApJ...785...74L, 2015ApJ...805...89L, 2016PhRvD..93d4065G}. GRB 211024B is likely such a case.

%If magnetars are the central engines of some GRBs, then several evolutionary results of magnetars are expected: (1) the immediate collapse into a BH; (2) the collapse of a supra- massive magnetar into a BH after it spins down; and (3) a stable magnetar \citep{2014ApJ...785...74L, 2015ApJ...805...89L, 2016PhRvD..93d4065G}. Our analysis suggests that some magnetars may indeed collapse into a BH, which provides direct support for the magnetar central engine model.

\subsubsection{Magnetar collapse caused by dipole radiation spin-down}\label{dipole_radiation}

A newborn magnetar experiences spin-down driven by both magnetic dipole radiation and gravitational wave emission \citep{2001ApJ...552L..35Z}. The spin-down rate is given by:
\begin{equation}
    \dot{E} = I\Omega\dot{\Omega} = -\frac{B_{\rm p}^2R^6\Omega^4}{6c^3} - \frac{32I^2\epsilon^2\Omega^6}{5c^5}
\end{equation}
where $I$ denotes the moment of inertia, $\Omega=2\pi/P$ denotes the angular frequency, $\dot{\Omega}$ denotes the time derivative thereof, $B_{\rm p}$ denotes the surface magnetic field strength at the poles, $R$ corresponds to the magnetar's radius, and $\epsilon$ is the ellipticity of the magnetar.

The spin-down rate can be dominated by either magnetic dipole radiation or gravitational wave radiation. When spin-down dominated by magnetic dipole radiation, 
\begin{equation}
    t_{\rm md} =  \frac{3c^3I}{B_{\rm p}^2 R^6 \Omega_0^2} \simeq 2.05\times 10^{3}\mbox{s }(I_{45} B^{-2}_{\rm p,15} P^{2}_{0,-3} R^{-6}_6),
\end{equation}
where $B_{\rm p,15}$ is the magnetic ﬁeld strength in units of $10^{15}$ G, $P_{0,-3}$ is the initial spin period of new-born magnetar in milliseconds, $R_6$ is magnetar radius in units of $10^6$ cm, $I_{45}$ is the moment of inertia in units of $10^{45}$g cm$^{-2}$. When spin-down dominated by gravitational wave emission,
\begin{equation}
    t_{\rm GW} = \frac{5c^5}{128GI\epsilon^2\Omega_0^4} \simeq 9.1 \times 10^3\mbox{s }(I_{45}^{-1} P_{0,-3}^4 \epsilon_{-3}^{-2}).
\end{equation}

The spin-down luminosity $L_0$ and the spin-down timescale $\tau$ of magnetar can be calculated by \citep{2001ApJ...552L..35Z},

\begin{equation}\label{eq.L0}
    L_0 = 1.0\times 10^{49}\mbox{ erg s}^{-1}\space(B^2_{\rm p,15}P^{-4}_{0,-3}R^6_6)
\end{equation}

\begin{equation}\label{eq.tau}
    \tau = \mbox{min} (t_{\rm md},t_{\rm GW})
\end{equation}

The relation of spin-down luminosity $L_0$ and the plateau luminosity $L_{X,iso}$ defind as 
\begin{equation}\label{eq.etaX}
    \eta_X L_0 = \eta_X(L_{\rm dis} + L_{\rm inj}) = L_{X,iso}f_b
\end{equation}
where $\eta_X$ is the radiation efﬁciency and $L_{\rm inj}$ is the energy injection luminosity.

As discussed in \cite{2001ApJ...552L..35Z, 2013ApJ...779L..25F, 2016MNRAS.458.1660L} and \cite{2022ApJ...934..125X}, if a newborn magnetar rotates rapidly ($P_0\sim 1$ ms) and has a significant ellipticity ($\epsilon > 10^{-3}$), the initial spin-down phase is dominated by gravitational wave emission. Otherwise, the spin-down is dominated by dipole radiation throughout. In a more general scenario, the spin-down mechanism would transition from gravitational wave dominated to magnetic dipole radiation dominated. For simplicity, we assume that the spin-down is dominated by only one of these mechanisms throughout the entire process.

% As discussed in \cite{2013ApJ...779L..25F, 2013ApJ...763L..22Z, 2015ApJ...805...89L} and \cite{2016PhRvD..93d4065G}, significant energy may be released in the form of GWs. For simplicity, we did not include the rotational energy loss of the supra-massive magnetar due to gravitational wave (GW) emission due to the uncertainty of the ellipticity of the magnetar. 

% The relation of spin-down luminosity $L_0$ and the plateau luminosity $L_{b,iso}$ defind as 
% \begin{equation}\label{eq.etaX}
%     \eta_X L_0 = L_{b,iso}f_b
% \end{equation}
% where $\eta_X$ is the radiation efﬁciency and $f_b=1-\cos{\theta_j}$ is the beaming factor.

The evolution of the spin period of magnetars is described by the following equation \citep{1983bhwd.book.....S}:
\begin{equation}
    P(t)=P_0\left(1+\frac{t}{\tau}\right)^{1/2}
\end{equation}

As the period of magnetar increase, the centrifugal force cannot resist gravity and collapses into a black hole, resulting a sharp decay in X-ray flux as observed.
The maximum gravitational mass of supramassive magnetars can be expressed as a function of the spin period \citep{2014PhRvD..89d7302L},
\begin{equation}\label{eq.mass_max}
    M_{\rm max} = M_{\rm TOV}(1+\hat{\alpha}P^{\hat{\beta}})
\end{equation}
where $M_{\rm TOV}$ is the maximum mass of non-rotating nuetron star (NS), $\hat{\alpha}$ and $\hat{\beta}$ depend on the NS EoS.

In this paper, we adopt neutron star EoS GM1 (the maximum mass of non-rotating NS is $M_{TOV}=2.37M_{\odot}$, the neutron star radius is $R=12.05$\,km, the moment of inertia is $I=3.33\times 10^{45}$g cm$^{-2}$, $\hat{\alpha}=1.58\times10^{-10}$\,s$^{-\hat{\beta}}$ and $\hat{\beta}=-2.84$) in the modeling. Recently, \cite{2016PhRvD..94h3010L} suggested that the “internal plateau” might be produced by a QS. Our model is not sensitive to the EoS, therefore our main results will still hold even if we adopt QS EoSs.

In Section \ref{sec.afterglow}, we derived the X-ray plateau flux $F_b\sim3.0\times 10^{-10}$ erg s$^{-1}$ cm$^{-2}$ and constrained GRB afterglow parameters under various assumptions. It is noteworthy that, the injection luminosity in the wind-like medium scenario is found to be remarkably high, reaching values of $10^{50}-10^{52}$ erg s$^{-1}$. Given that the rotational energy of a typical magnetar is of the order of $10^{52}$ erg, it is evident that additional energy sources must be invoked to sustain such high luminosity. Fallback accretion-powered jets represent a plausible alternative. Here, we only consider the case of k=0 with a fixed $t_e$, where $\tau \sim t_e = t_{\rm col} = t_b/(1+z)=7.10\times 10^3$\,s. By substituting these afterglow parameters into Equations (\ref{eq.L0}), (\ref{eq.tau}) and (\ref{eq.etaX}), we obtain the magnetar's initial spin period ($P_0$) and magnetic field ($B_p$) for both dipole radiation-dominated and gravitational wave-dominated regimes. For the dipole-dominated case, $P_0=8.4$\,ms and $B_p=4.7\times10^{15}$\,G. For the gravitational wave-dominated case, $P_0=0.79$\,ms, $B_p=3.3\times10^{14}$\,G. 
The magnetar's radiative efficiency ($\eta_X$) is found to be 0.07. It is worth noting that the spin-down timescale typically exceeds the collapse time. Consequently, the derived spin period and magnetic field strength should be interpreted as upper limits.

\begin{figure*}[ht!]
\plotone{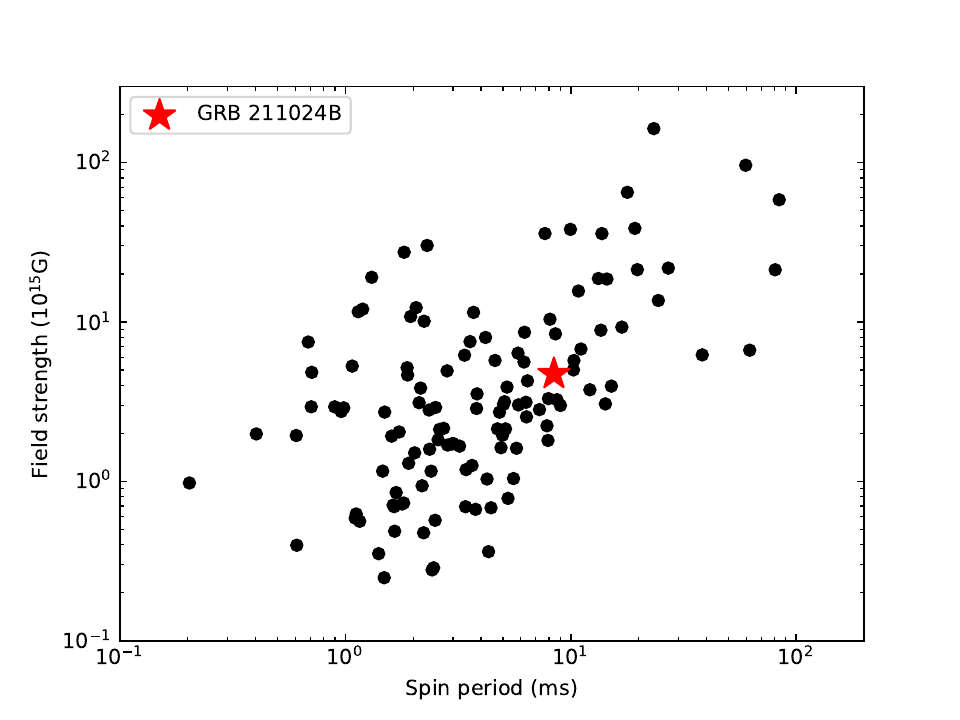}
\caption{Magnetic ﬁeld strengths versus spin periods of magetars for \textit{Swift} GRB samples \citep[black dot, from][]{2014MNRAS.443.1779R}. The magnetar corresponding to GRB 211024B is marked as red star. It is important to note that all data points in the figure are derived under the assumption of EM-dominated spin-down. \label{fig.BP}}
\end{figure*}

\cite{2014MNRAS.443.1779R} compiled a sample of \textit{Swift} X-ray afterglow light curves with known redshifts from January 2005 to December 2013. They modeled these light curves using a magnetar model, assuming energy release via dipole radiation, deriving magnetar spin periods ranging from $0.2$\,ms to $100$\,ms and magnetic field strengths between $2.0\times 10^{14}$\,G and $2.0\times 10^{17}$\,G (represented by black dots in Figure \ref{fig.BP}). Since spin-down luminosity $L_0$ and spin-down time scale are estimated from observational data, it implies that their uncertainties are difficult to estimate. Consequently, the uncertainties of the derived parameters, namely the spin period and magnetic field strength, are also difficult to estimate. The data points in Figure \ref{fig.BP} therefore do not include error bars. The magnetar associated with GRB 211024B (marked as stars in Figure \ref{fig.BP}) exhibits typical values of the whole sample.

%A pair shock were generate, namely revers shock (RS) and forward shock (FS). The RS propagate into the shell and an FS propagate into the ISM \citep{Rees1992,Meszaros1997,Sari1998, Sari1999,Zou2005}. 

\subsubsection{Magnetar collapse caused by accretion}

Following a supernova explosion, a significant amount of material is ejected into space at extremely high velocities. However, a fraction of this material fails to achieve escape velocity and subsequently falls back onto the central remnant. In such cases, if the central remnant is a magnetar, the evolution of spin and mass deviate from the conventional models presented in Section \ref{dipole_radiation}. \cite{2012ApJ...759...58D} conducted a comprehensive analysis of the influence of hyperaccreting fallback disks on the spin evolution of millisecond magnetars. In this case, the spin evolution of a magnetar is governed by two primary factors: magnetic dipole radiation, and hyperaccreting fallback disks. The spin evolution of magnetar can be given by
\begin{equation}
    \frac{\mbox{d}(I\Omega_s)}{\mbox{d}t} = \tau_{\rm acc} + \tau_{\rm dip}
\end{equation}
where $I$ is the stellar moment of inertia, $\tau_{\rm acc}$ and $\tau_{\rm dip}$ are net torque exerted on the magnetar by the accretion disk and the torque due to magnetic dipole radiation, respectively \citep[the specific form is detailed in ][]{2012ApJ...759...58D}. The early spin evolution of a magnetar is primarily influenced by accretion, particularly for low initial spin rates or high accretion rates, resulting in spin-up. In the late stages of evolution, however, magnetic dipole radiation becomes dominant, causing the magnetar to spin-down. Under the dual effects of accretion and spin-down, if the magnetar's initial mass is sufficiently large or it accretes enough matter, the magnetar's mass may exceed its maximum gravitational mass, causing it to collapse into a black hole. Before collapse, the dipole radiation of magnetar is
\begin{equation}\label{eq.Ldip}
    L_{\rm dip} = 9.6\times10^{48}\rm{erg\ s}^{-1}\sin^2\chi \left(\frac{B_0R^3}{10^{33} \rm{G\ cm}^3}\right)^2\left(\frac{P}{1\rm{ms}}\right)^{-4}
\end{equation}
where $\chi$ is the inclination angle of the magnetic axis to the spin axis, $R$ is the radius of megnetar, $B_0$ is the surface magnetic field strength, and $P=2\pi/\Omega_s$ is the time-dependence spin period. The parameterize fallback accretion rate can be described by \citep{2001ApJ...550..410M, 2008ApJ...683..329Z}
\begin{equation}
    \dot{M} = 10^{-3}\left[(\eta t^{1/2)})^{-1} + (\eta t_p^{13/6} t^{-5/3})^{-1}\right]^{-1}M_{\odot} \mbox{ s}^{-1}
\end{equation}
where $\eta$ is a factor that accounts for explosion energy, and $t_p$ is the time of peek accretion rate. For more energetic explosion, $\eta$ is smaller and $t_p$ is larger. Therefore, the baryonic mass of the magnetar at time $t$ can be determined as 
\begin{equation}
    M_b(t) = M_0+\int_{t_0}^t\dot{M}\mbox{d}t
\end{equation}
where $M_0$ is the initial mass and $t_0$ is the start time of fall-back accretion. Considering a portion of the accreted matter is radiated away in the form of neutrinos, the corresponding gravitational mass is calculated to be
\begin{equation}
    M = M_b(t)\left[1+\frac{3}{5}\frac{GM_b(t)}{Rc^2}\right].
\end{equation}

\begin{figure*}[ht!]
\gridline{\fig{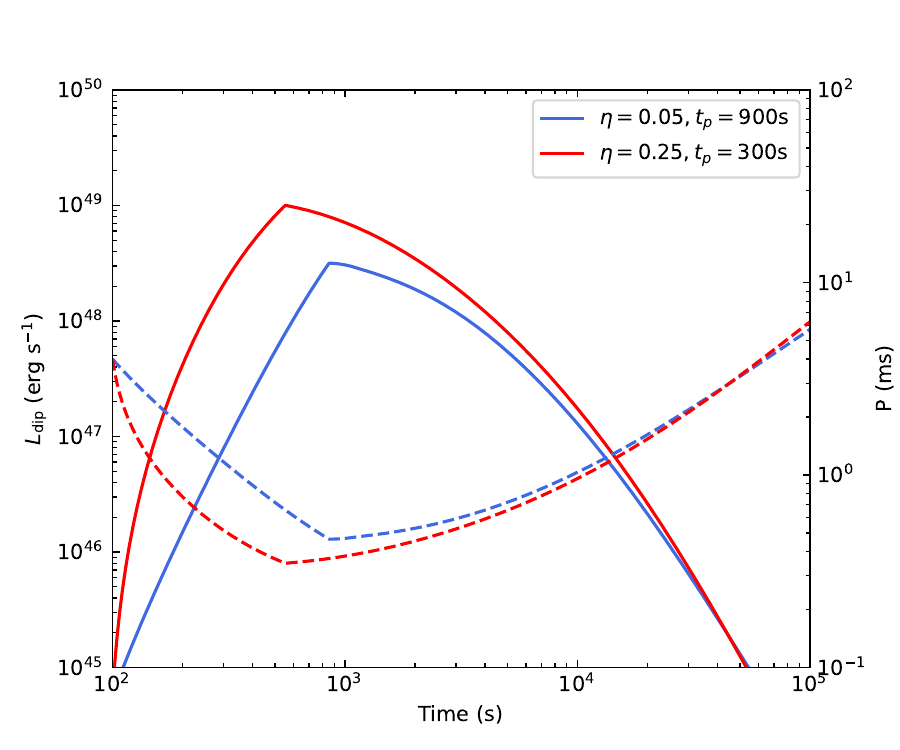}{0.48\textwidth}{(a)}
          \fig{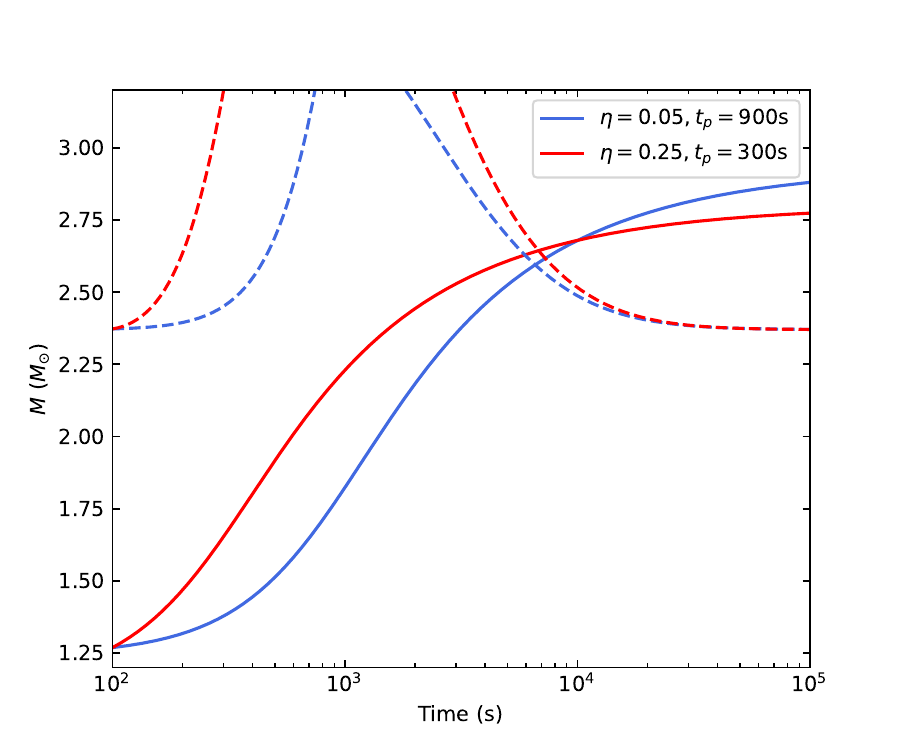}{0.48\textwidth}{(b)}
          }
\caption{(a) Temporal evolution of magnetic dipole luminosity (solid line) and spin period (dashed line); (b) temporal evolution of gravitational mass (solid line) and maximum gravitational mass (dashed line, described by Equation \ref{eq.mass_max}). The blue lines correspond to case of $\eta=0.05$, $t_p=900$ s, and the red lines correspond to case of $\eta=0.25$, $t_p=300$ s. \label{fig.hyperacc}}
\end{figure*}

Through numerical integration of differential Equation \ref{eq.Ldip}, we can determine the evolutionary of the spin and luminosity of magnetar. We adopt an initial mass of $M_0=1.4\ M_{\odot}$, an initial radius of $R_0=12.05$ km, an initial spin period of $P_0=4$ ms, a surface magnetic field of $B_0=10^{14}$ G, accretion start time of $t_0=100$ s, and $\sin^2\chi = 0.5$ for the magnetar. Figure \ref{fig.hyperacc} presents these evolutionary curves for two specific cases: $\eta=0.05$, $t_p=900$ s, and $\eta=0.25$, $t_p=300$ s. For both scenarios, the magnetar exhibits an initial spin-up phase within the first 1 ks, followed by a subsequent spin-down phase dominated by dipole radiation. Eventually, the magnetar accretes sufficient matter to exceed the maximum gravitational mass, resulting in a collapse into a black hole around 7 ks (source frame time). The pre-collapse dipole radiation luminosity is estimated to be $\sim 4\times 10^{47}$ erg s$^{-1}$, consistent with observation results.

It is worth noting that Equation \ref{eq.Ldip} does not account for gravitational wave radiation. In fact, when the magnetar's spin period accelerates to 1 ms, gravitational wave radiation may become non-negligible \citep{2001ApJ...552L..35Z}. At this point, spin-up becomes difficult, and the period evolution enters a brief plateau phase, leading to a plateau in the radiation luminosity as well. As the spin-down dominates at later times, both gravitational wave radiation and accretion effects become negligible.

Studies of the X-ray light curves of short gamma-ray bursts have shown \cite[e.g.,][]{2013MNRAS.430.1061R,2020PhRvD.101f3021S} that these rapidly rotating magnetars, born from binary mergers, collapse due to spin-down with collapse timescales ranging from $\sim 10^2$ s to $\sim 10^3$ s. This statistical result differs from our derived collapse timescale of $t_{col} \sim 7$ ks, which suggests that the collapse mechanism of the magnetar responsible for GRB 211024B may differ from the standard spin-down scenario. Following a supernova explosion, a fraction of the material that fails to escape the gravitational potential well and eventually falls back onto the central object, increasing its mass. In such cases, if the central object is a magnetar, its collapse would be induced by accretion. Furthermore, the torque from an early accretion disk can accelerate the magnetar's spin initially, followed by spin-down due to dipole radiation, which can extend the magnetar's collapse timescale. Both of these factors can contribute to the observed long collapse timescales.

\subsection{The origin of early optical emission}
A subset of GRB early optical afterglow observations exhibit light curve flux profiles characterized by an initial rise followed by a decline. This feature can be attributed to either a forward or reverse shock. If the emission is attributed to a forward shock, the time corresponding to the peak flux represents the deceleration timescale of the jet, which can be used to constrain the initial Lorentz factor. If the emission is attributed to a reverse shock, the time corresponding to the peak flux represents the crossing timescale of the reverse shock \citep{Rees1992,Meszaros1997,Sari1998, Sari1999,Zou2005}. \cite{2003ApJ...597..455K} studied in detail the light curve features of reverse shocks in the thin-shell and thick-shell cases. Under typical afterglow parameters, for the ISM, the typical rise and decay indices of the light curve are $5$ or $1/2$ and $-2$, respectively; for the Wind model, the typical rise and decay indices of the light curve are $5/2$ or $1/2$ and -3, respectively \citep[see Figure 1 in][]{2003ApJ...597..455K}. In Section \ref{subsec.temporal}, we obtain rise and decay indices of $1.1$ and $-0.96$ for the early optical emission of GRB 211024B, which are inconsistent with theoretical expectations. Of course, it is also possible that its afterglow parameters are rather special. On the other hand, the early optical emission could be the result of a combined contribution from a reverse shock and a forward shock \citep{2023ApJ...948...30Z}. The presence of a extra component can lead to the fitted decay index not accurately reflecting the emission characteristics of the reverse shock. Due to the lack of earlier and multi-band observational data, we cannot further explore these cases.

\section{Conclusion}
We present optical follow-up observations of ultra-long GRB 211024B from $\sim 200$\,seconds to $\sim 65$\,days. The early phase X-ray light curve exhibits a two-part structure characterized by an initial, slowly declining plateau followed by a very steep decay, which is a typical ``internal plateau" feature. This is often taken as evidence for a magnetar as the central engine. In the optical r-band, the early bump-like light curve flux is lower than the extrapolation from the late-time afterglow. To account for the observed increase in late-time afterglow flux, we propose an energy injection mechanism into the external shock. We incorporated an energy injection term into our afterglow fitting, finding injected luminosities ranging from $10^{47}$ to $10^{52}$ erg s$^{-1}$ depending on the circumburst medium. For the ISM with a fixed $t_e$ scenario, energy injection from magnetar dipole radiation provides a plausible explanation, with derived magnetar parameters consistent with the general magnetar population. However, for the other three scenarios, the energy injection termination time or luminosity deviates from magnetar model predictions, suggesting alternative energy sources.

If the magnetar is surrounded by a hyperaccretion disk, neutrino annihilation and magnetic activity from the stellar surface could produce relativistic jets along the magnetic poles \citep{2009ApJ...703..461Z, 2010ApJ...718..841Z}, injecting energy into the external shock. Additionally, fallback accretion onto a black hole formed from magnetar collapse could generate relativistic jets through the Blandford-Znajek mechanism \citep{Blandford1977}. These processes could explain the observed discrepancies between the energy injection parameters and the magnetar dipole radiation model.

The possibility that the early bump-like feature is caused by a reverse shock cannot be ruled out. In this case, the entire optical light curve can be explained without the need for a magnetar energy injection model. Unfortunately, due to the lack of early and multi-band observational data, we are unable to confirm whether this is the case with GRB 211024B. For the current observational data, the most natural explanation is still the magnetar model discussed in this paper.

The case of GRB 211024B, along with previously studied ULGRBs (e.g., GRB 101225A, GRB 111209A), highlights the crucial role of an active central engine in these events. Multi-band observations can reveal the signatures of the central engine, enabling us to improve our understanding of ULGRBs and constrain the physical parameters of the engine itself.

\begin{acknowledgments}
% NOT
The data presented here were obtained in part with ALFOSC, which is provided by the Instituto de Astrofisica de Andalucia (IAA) under a joint agreement with the University of Copenhagen and NOT. This research has made use of the Spanish Virtual Observatory (http://svo.cab.inta-csic.es) supported by the MINECO/FEDER through grant AyA2017-84089.7. 
% Swift
This work made use of data supplied by the UK \textit{Swift} Science Data Centre at the University of Leicester. 
D.X. acknowledges the science research grants from the China Manned Space Project with NO. CMS-CSST-2021-A13 and CMS-CSST-2021-B11. 
This work is supported by the National Natural Science Foundation of China under grant 12473012.
\end{acknowledgments}

\vspace{5mm}
\facilities{NEXT, NOT(ALFOSC), VLT(X-Shooter), LBT, \textit{Swift}(BAT, XRT and UVOT)}

\software{Astropy \citep{2013A&A...558A..33A,2018AJ....156..123A},  
          Source Extractor \citep{1996A&AS..117..393B},
          emcee \citep{2013PASP..125..306F}, 
          IRAF \citep{1986SPIE..627..733T}
          }

% \appendix

% \section{Appendix information}

\bibliography{citation}{}
\bibliographystyle{aasjournal}

\end{document}